\documentclass[12pt,twoside]{article}   
\usepackage[super,sort,comma]{natbib}

\usepackage{fancyhdr}		

\usepackage[section]{placeins}   %

\makeatletter \renewcommand\@biblabel[1]{$^{#1}$} \makeatother
\newcommand{\cen}[1]{\begin{center} #1 \end{center}}

\usepackage[]{lineno}\relax

\usepackage{setspace}
\usepackage{amssymb}
\usepackage{amsmath}
\usepackage{multirow}
\usepackage{booktabs}
\usepackage{dcolumn}
\usepackage{enumerate}
\usepackage{authblk}
\usepackage{makecell}
\usepackage[font=small,labelfont=bf]{caption}
\usepackage[colorlinks=true,linkcolor=blue,urlcolor=black,bookmarksopen=true,citecolor=red,bookmarks=true]{hyperref}

\usepackage[labelformat=simple]{subcaption}

\usepackage{graphicx}
\usepackage[para,online,flushleft]{threeparttable}

\let\oldequation\equation
\let\oldendequation\endequation
\renewenvironment{equation}{\linenomathNonumbers\oldequation}{\oldendequation\endlinenomath}

\begin{document}

\cen{\sf {\Large {\bfseries Generalized-Equiangular Geometry CT: Concept and Shift-Invariant FBP Algorithms} \\  
	\vspace*{10mm}
	Yingxian~Xia, Zhiqiang~Chen, Li~Zhang, Yuxiang~Xing and Hewei~Gao$^{1,2*}$}\\
	\vspace*{5mm}
	$^{1}$Department of Engineering Physics, Tsinghua University, Beijing 100084, China\\
	$^{2}$Key Laboratory of Particle \& Radiation Imaging (Tsinghua University), Ministry of Education, China\\
}

\setcounter{page}{1}
\pagestyle{plain}
*Author to whom correspondence should be addressed. Email: hwgao@tsinghua.edu.cn \\
\begin{abstract}
\textbf{Background}: 
With advanced X-ray source and detector technologies being continuously developed, non-traditional CT geometries have been widely explored. Generalized-Equiangular Geometry CT (GEGCT) architecture, in which an X-ray source might be positioned radially far away from the focus of arced detector array that is equiangularly spaced, is of importance in many novel CT systems and designs.

\textbf{Purpose}:  
GEGCT, unfortunately, has no theoretically exact and shift-invariant analytical image reconstruction algorithm in general. In this study, to obtain fast and accurate reconstruction from GEGCT and to promote its system design and optimization, an in-depth investigation on a group of approximate Filtered BackProjection (FBP) algorithms with a variety of weighting strategies has been conducted.

\textbf{Methods}:
The architecture of GEGCT is first presented and characterized by using a normalized-radial-offset distance (NROD). Next, shift-invariant weighted FBP-type algorithms are derived in a unified framework, with pre-filtering, filtering, and post-filtering weights. Three viable weighting strategies are then presented including a classic one developed by Besson in the literature and two new ones generated from a curvature fitting and from an empirical formula, where all of the three weights can be expressed as certain functions of NROD. After that, an analysis of reconstruction accuracy is conducted with a wide range of NROD. We further stretch the weighted FBP-type algorithms to GEGCT with dynamic NROD. Finally, the weighted FBP algorithm for GEGCT is extended to a three-dimensional form in the case of cone-beam scan with a cylindrical detector array.

\textbf{Results}:
Theoretical analysis and numerical study show that weights in the shift-invariant FBP algorithms can guarantee highly accurate reconstruction for GEGCT.
A simulation of Shepp-Logan phantom and a GEGCT scan of lung mimicked by using a clinical lung CT dataset both demonstrate that FBP reconstructions with Besson and polynomial weights can achieve excellent image quality, with Peak Signal to Noise Ratio and Structural Similarity being at the same level as that from the standard equiangular fan-beam CT scan.
Reconstruction of a cylinder object with multiple contrasts from simulated GEGCT scan with dynamic NROD is highly consistent with fixed ones when using the Besson and polynomial weights, demonstrating the robustness and flexibility of the presented FBP algorithms.
In terms of resolution, the direct FBP methods for GEGCT could achieve $1.35$ lp/mm of spatial resolution at $10\%$ modulation transfer functions point, higher than that of the rebinning method which can only reach $1.14$ lp/mm.
Moreover, 3D reconstructions of a disc phantom reveal that a greater value of NROD for GEGCT will bring less cone beam artifacts as expected.

\textbf{Conclusions}:
We propose the concept of GEGCT and investigate the feasibility of using shift-invariant weighted FBP-type algorithms for reconstruction from GEGCT data without rebinning. A comprehensive analysis and phantom studies have been conducted to validate the effectiveness of proposed weighting strategies in a wide range of NROD for GEGCT with fixed and dynamic NROD.

\vspace{5mm}
{\bf Keywords}: Equiangular CT, Filtered Backprojection, Weighted FBP, Stationary CT.

\end{abstract}

\newpage
\setlength{\baselineskip}{0.7cm}      
\pagestyle{fancy}

\nolinenumberdisplaymath
\section{Introduction} 
As X-ray source and detector technologies keep evolving, conventional and non-conventional Computed Tomography (CT) geometries have been continuously conceived and explored, aiming at faster scanning, better image quality, lower radiation dose, or more applications. \cite{RN183, RN186, RN187}
Historically, conventional CT has been developed for at least five generations.\cite{CTgene}
In this being, the third-generation CT, using a substantially larger fan-beam coverage and a correspondingly longer detector array, has become the most commonly used CT geometry.
Meanwhile, over years non-conventional CT architectures draw great interest in the literature as well. Multi-source CT, inverse-geometry CT (IGCT), stationary CT, and many others are under active investigations.

The first multi-source CT system was commercialized in 2006, which has a dual-source helical CT system design equipped with two X-ray tubes and two corresponding multi-slice detectors, mounted onto the rotating gantry with an angular offset of $90^o$. \cite{flohr2006first}
Furthermore, a dedicated cardiovascular dual-source CT system was introduced in 2017, in which two X-ray sources are used to enable dual overlapping cone beams to achieve 140 mm coverage at a rotation speed of 0.24 sec. 
Chen et al. proposed a triple-source CT system with three pairs of source and flat panel detectors in 2013. \cite{chen2013reconstruction}
Besson also introduced a novel triple-source CT system with two independently rotating drums, supporting three X-ray sources and an angularly extended detector arc separately. \cite{besson2015new}
Besides, several multi-source and triple-source CT architectures for cardiac CT scanning were proposed in 2017. \cite{FPEG2017}
Beyond dual-source or triple-source, Wu et al. developed a new swinging multi-source industrial CT system. \cite{wu2017swinging}
Besides multi-source CT systems, IGCT has also been actively investigated. The idea of IGCT was firstly reported in 1993 to solve the data insufficiency problem and eliminate cone-beam artifacts. \cite{1993Large}
Later, Schmidt et al. in 2004 studied the feasibility of an inverse-geometry volumetric CT system using a large-area scanning source, \cite{RN122} and provided a table-top prototype inverse-geometry volumetric CT system in 2006. \cite{RN124} 

Given the concepts of multi-source CT and IGCT, CT architectures with stationary sources were proposed naturally.
Hsieh et al. presented a stationary source IGCT system comprised of circumferential X-ray source arrays to achieve better temporal resolution in 2013. \cite{RN129}
Chen et al. showed a concept of helical interlaced source detector array CT in 2014 which could complete axial coverage quickly. \cite{RN119}
In 2018, Cramer et al. designed a CT system with a ring of multiple miniature x-ray sources for CT imaging without any moving parts. \cite{cramer2018stationary}
Besides the stationary CT with circular arrangements, researchers also explored the feasibility of CT imaging with straight line trajectory.\cite{smith1993fan,sidky2005volume,gao2007direct,gao_straight-line-trajectory-based_2013}
In 2020, Zhang et al. proposed a symmetric-geometry computed tomography, where the x-ray source and detector are linearly distributed in a symmetric and stationary design. \cite{RN133,RN134}
Luo et al. evaluated simulation results of potential system geometry for a stationary head CT using a set of linearly distributed carbon nanotube x-ray sources in 2021. \cite{RN117}

\begin{figure*}
	\centering
	\begin{subfigure}[b]{0.3\textwidth}
		\centering
		\includegraphics[width=\textwidth]{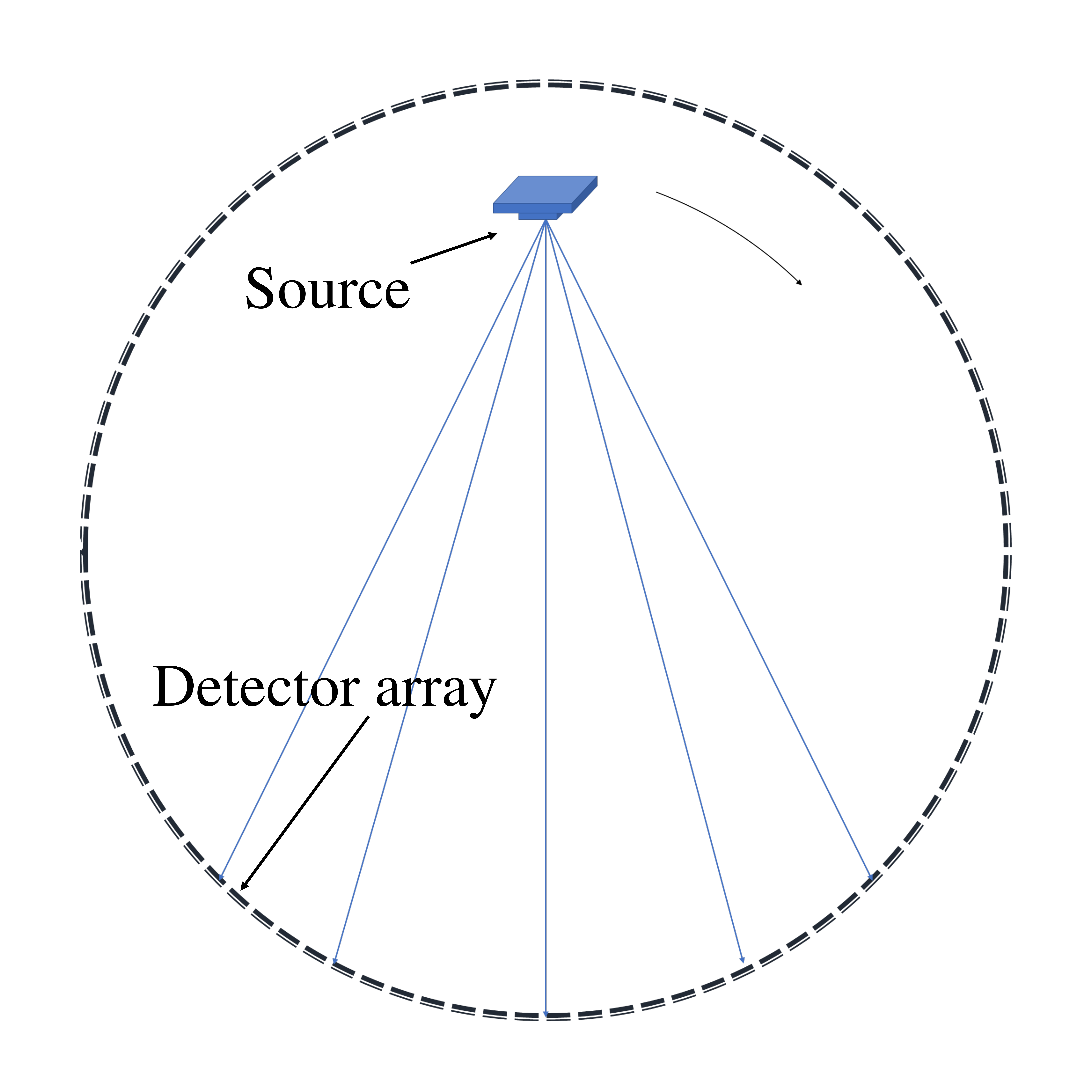}
		\caption{}
		\label{fig:applyCT1}
	\end{subfigure}
	\begin{subfigure}[b]{0.3\textwidth}
		\centering
		\includegraphics[width=\textwidth]{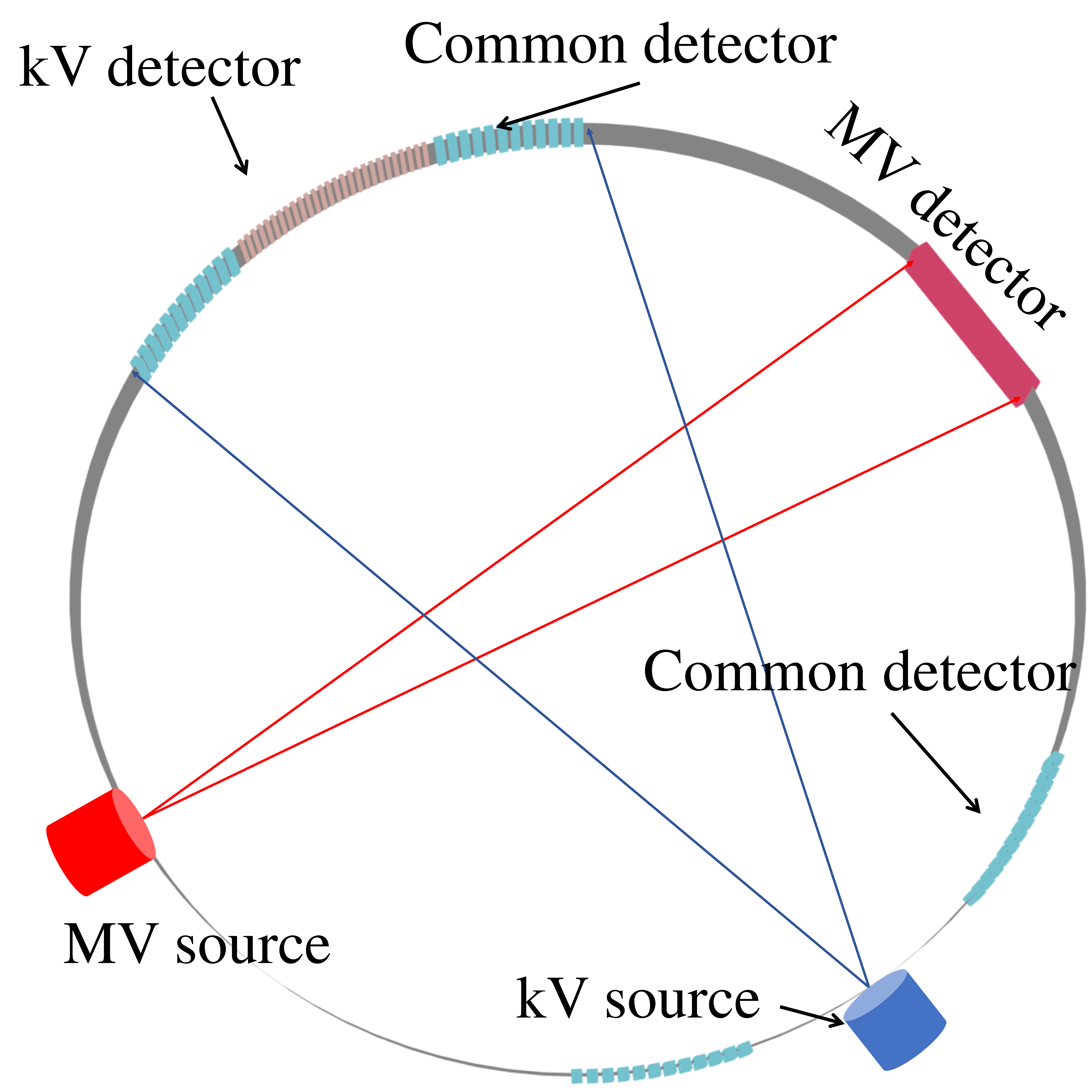}
		\caption{}
		\label{fig:applyCT2}
	\end{subfigure}
	\begin{subfigure}[b]{0.3\textwidth}
		\centering
		\includegraphics[width=\textwidth]{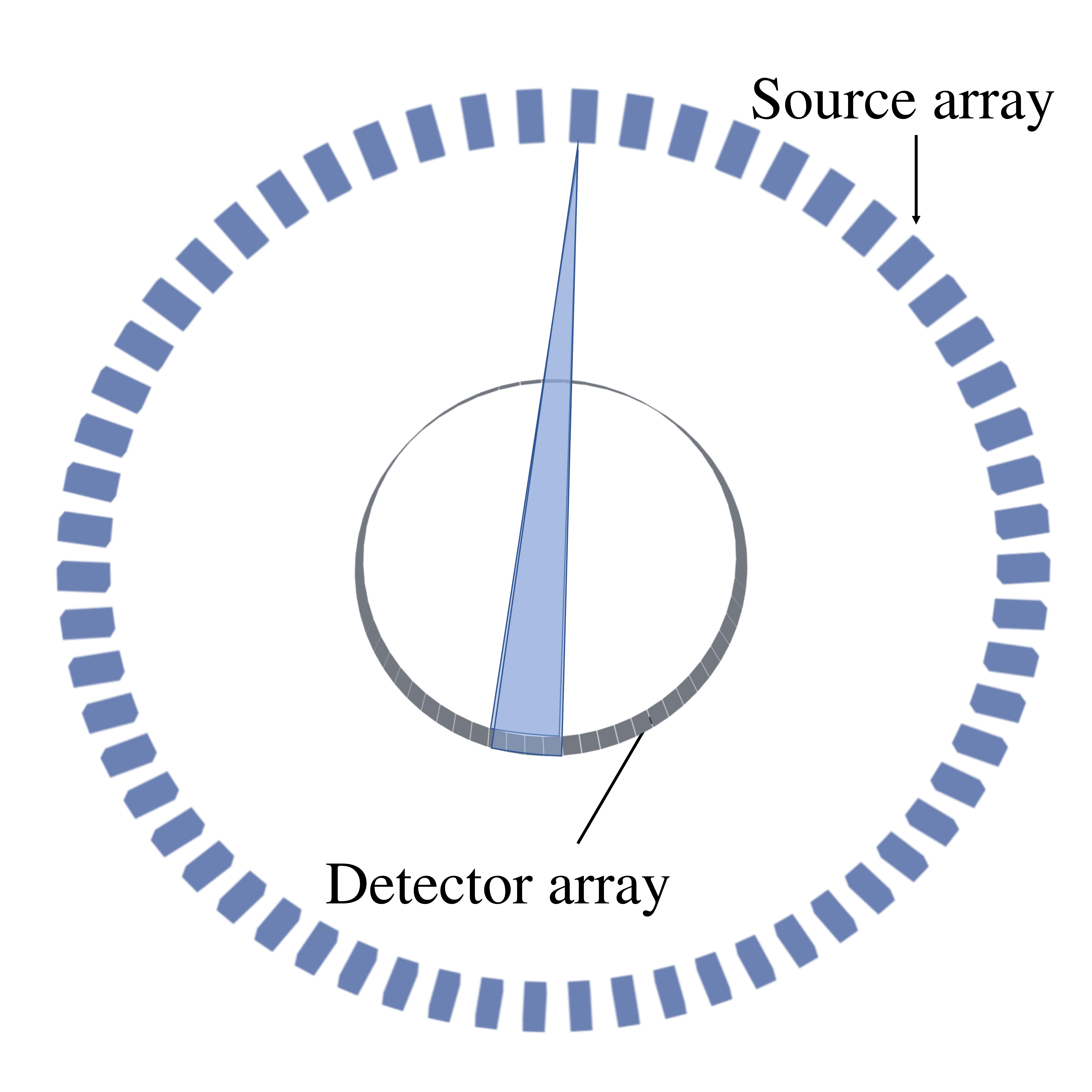}
		\caption{}
		\label{fig:applyCT3}
	\end{subfigure}
	\caption
	{ Exemplary applications of GEGCT architecture.
		(a) The fourth generation CT with an X-ray source rotating and a ring of detector arrays;
		(b) A design of compact architecture of co-planar transmission and emission guidance in radiotherapy (Co-TeGRT);
		(c) A stationary CT with a ring of X-ray sources and a ring of detector arrays, which might exist z-shift between rings. \cite{wu2021stationary, tucker2020design}}
	\label{fig:applyCT} 
\end{figure*}

In this work, we observe a new and interesting CT geometry with favorable potential in real applications, when exploring the ring-based stationary CT and comparing it with the existing fourth generation CT and the newly proposed CO-planar Transmission and Emission Guidance for Radiotherapy \cite{gao2020novel}(as shown in the Fig.~\ref{fig:applyCT}).
Specifically, in these CT system design, the X-ray sources are seen to be positioned all along the radial line made by the center and the focus of an arc-shaped detector array. 
For such a system, although the arced detector is kind of equiangularly distributed, the corresponding CT scan does not follow a truly equiangular data sampling pattern except for two particular positions. As a result, it is called Generalized-Equiangular Geometry CT (GEGCT) in this paper.

Typically, regardless of vast variabilities in conventional and non-conventional CT geometries, fan-beam CT detector shares only three kinds of arrangements, arc- or ring-shaped, flat-shaped, or a hybrid of the two, leading to equiangular or equispaced data sampling on the detector globally or locally. Wei et al. developed a general formula for fan-beam CT in 2005.\cite{2005General} And it is well-known that image reconstruction from a standard equiangular or equispaced fan-beam CT can be done efficiently and exactly by using the classic Filtered BackProjection (FBP) algorithms.

For GEGCT, although CT projection data is collected on a regular arc-shaped detector, the transfer from the fan angle with respect to the focus to the fan angle with respect to the source is nonlinear and could vary with the scan view angle as well.
As a result, projection ray angles are not indeed uniformly sampled, leading to the loss of shift-invariant in the filtering step in the FBP reconstruction.
On the other hand, shift-invariant filtering is critical to the computational efficiency for FBP, as its implementation can be realized by the Fast Fourier Transform (FFT) technique. As a matter of fact, keeping the FBP algorithm shift-invariant property without degrading image quality, in general, is an important goal when developing image reconstruction for a specific CT system.

In the literature, Besson studied the general parametrizations of fan-beam CT that could have direct FBP method without rebinning.\cite{RN181}
Kachelriess also proposed several interesting detector shapes as well for compact CT systems, which could perform FBP in the native geometry without rebinning. \cite{RN155}
Different from the above, Pan et al. proposed a fast reconstruction with shift-variant filtration for fan-beam CT by avoiding calculating the spatially variant weighting factor $L$. \cite{RN165} 

Although various CT geometries related to GEGCT have been explored before, \cite{RN129, RN119,cramer2018stationary} to the best of our knowledge, there is still lack of a systematical study of this special imaging geometry.
Besson's study on CT fan-beam parametrizations led to a weighted FBP-type reconstruction that fits the fourth generation CT in 1996, \cite{RN181} equivalent to GEGCT with normalized-radial-offset-distance (NROD) close to $1$.
It is still desired to develop and conduct an in-depth analysis of approximate FBP algorithms for GEGCT that could work efficiently for a wide range of NROD (both fixed and dynamic), where an appropriate weighting is critical to the reconstruction accuracy.

In this paper, we first propose the concept of GEGCT and characterize its imaging geometry by using NROD.
Shift-invariant weighted FBP-type algorithms are then derived in a unified framework, with pre-filtering, filtering, and post-filtering weight, respectively. Three viable weighting strategies are presented where all of them can be expressed as certain functions of NROD. 
A comprehensive phantom simulation study along with a mimicked GEGCT scan of lung by using a clinical lung CT dataset is conducted to validate that highly accurate reconstruction can be guaranteed in a wide range of NROD. A case of using dynamic NROD during a GEGCT scan further demonstrates the effectiveness of the presented shift-invariant FBP algorithms for GEGCT, which could be leveraged to develop more flexible CT systems for practical applications.   
\section{Materials and Methods}
\subsection{The Concept of GEGCT}
\begin{figure}[htb]
	\centering
	\includegraphics[width=0.5\textwidth]{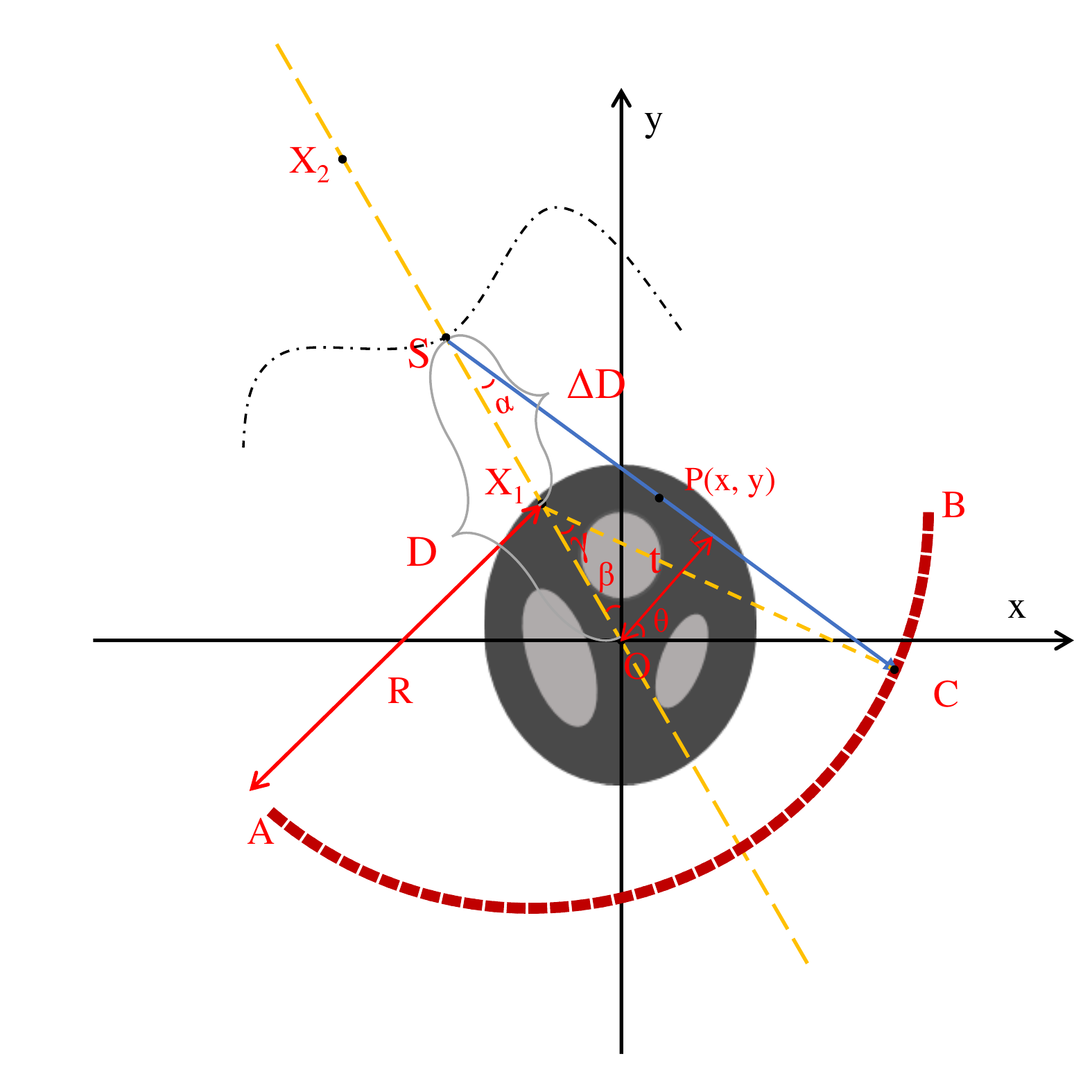}
	\caption{
		A schematic illustration of GEGCT geometry, where the arc AB represents the detector arrays with their nominal focus pointing to $X_1$ and radius denoted as $R$; the fan angle with respect to the source is denoted as $\alpha$, while the fan angle with respect to the detector focus is denoted as $\gamma$, and CT scan view angle as $\beta$.
		With the distance from the X-ray source to the isocenter (O) and the detector focus denoted as $D$ and $\Delta{D}$, respectively, the distance from the detector center to the isocenter can be denoted as $\mathrm{DID} = R - D + \Delta{D}$.}
	\label{fig:sketch} 
\end{figure}

Same as in the standard equiangular CT, an arc-shaped detector array is also employed in GEGCT. 
The difference, however, is that the X-ray source in GEGCT can be positioned radially far away from the focus of the arced detector array as shown in Fig.~\ref{fig:sketch}.
In order to characterize how far away X-ray source deviate from the focus of detector array, we compute the NROD and use it as a key parameter of differentiator for GEGCT, which is defined by the ratio of distance between the source and the detector focus to the radius of the detector arc,
\begin{equation}
k = \frac{\Delta D}{R}.
\end{equation}

As illustrated in Fig.~\ref{fig:sketch}, the fan angle of a specific detector element with respect to the source, $\alpha$, can be written as a function of NROD ($k$) and the fan angle of detector with respect to the focus of the detector, $\gamma$,
\begin{equation}
	\alpha_k (\gamma) =  \arctan \left(\frac{\sin(\gamma)}{\cos(\gamma)+k}\right) .
	\label{trans_akgama}
\end{equation}
At a specific view angle $\beta$ and detector fan angle $\gamma$, one can compute the distance from the isocenter to the projection ray $SC$, $t = D \sin \alpha_k(\gamma)$, and the projection ray angle $\theta = \beta+\alpha_k(\gamma)$. Therefore, the projection from GEGCT can be expressed as
\begin{equation}
	\begin{split}
		p_k (\gamma,\beta) 
		\equiv & \,\,\, p(\alpha_k (\gamma),\beta) \\
		= &\iint_{R^2} \mathrm{d} x \mathrm{d} y f(x,y) \delta [ x\cos(\beta+\alpha_k (\gamma)) + y\sin(\beta+\alpha_k (\gamma)) - D \sin \alpha_k (\gamma) ] \\
		= &\iint_{R^2} \mathrm{d} x \mathrm{d} y f(x,y) \frac {\delta[\sin(\alpha_k(\gamma) - \alpha_0)]}{L}  
	\end{split}
	\label{eqline}
\end{equation}
where,
\begin{equation}
\alpha_0 = \arctan \left( \frac{y\sin\beta+x\cos\beta}{D-y\cos\beta+x\sin\beta} \right)
\nonumber
\end{equation}
is the fan angle for the X-ray passing through the point $P(x, y)$, and
\begin{equation}
L = \sqrt{(D-y\cos\beta+x\sin\beta)^2 + (y\sin\beta+x\cos\beta)^2},\nonumber
\end{equation}
is the length of $SP$ in the Fig.~\ref{fig:sketch}.
For the projection from GEGCT with dynamic $k$, $\alpha_k(\gamma)$ will be a function of view angle $\beta$ as $k$ changes with $\beta$.

Comparing with the standard equiangular fan-beam CT, the occurrence of $k$ in Eq.~(\ref{eqline}) complicates the line integral formula and brings special behaviors for GEGCT. As $k$ changes from negative to greater than one, five different scenarios for $k$ will happen as follows.
\begin{itemize}
	\item[(\uppercase\expandafter{\romannumeral1})] $k<0$: in this case, the source-detector distance is less than the radius of the detector $R$. The standard equispaced fan-beam CT could be considered as a special case of GEGCT with $R \to \infty $, resulting in
	\begin{equation}
	k = \frac{DID+D-R}{R} \to -1.
	\nonumber
	\end{equation}
	\item[(\uppercase\expandafter{\romannumeral2})] $k=0$: in this particular case, the X-ray source is at the isocenter, so GEGCT is degenerating to the traditional equiangular fan-beam CT in terms of data sampling. Therefore, it has theoretically exact and shift-invariant analytical reconstruction algorithms.
	\item[(\uppercase\expandafter{\romannumeral3})] $0<k<1$: in this case, the source-detector distance is greater than $R$. Different detector elements will have different central angles with respect to the X-ray source, leading to a non-equiangularly sampled CT scan. An exemplary application of $0<k<1$ is the typical fourth generation CT as shown in Fig.~\ref{fig:applyCT1}.
	\item[(\uppercase\expandafter{\romannumeral4})] $k=1$: in this particular case, the X-ray source is positioned at the same circle with the detector ring. From the circumferential angle theorem, we know that all detector elements have the same central angles, again, with respect to the X-ray source. One might see this kind of imaging geometry in the Co-TeGRT design, as shown in Fig.~\ref{fig:applyCT2}.
	\item[(\uppercase\expandafter{\romannumeral5})] $k>1$: in this case, distance between the X-ray source and the focus of the detector ring is greater than the radius. Again, different detector elements will have different central angles with respect to the X-ray source. This kind of imaging geometry could be seen in the stationary CT with ring- or arc-shaped detector design as illustrated in Fig.~\ref{fig:applyCT3}.
\end{itemize}

\begin{figure}[htb]
	\centering
	\includegraphics[width=0.45\textwidth]{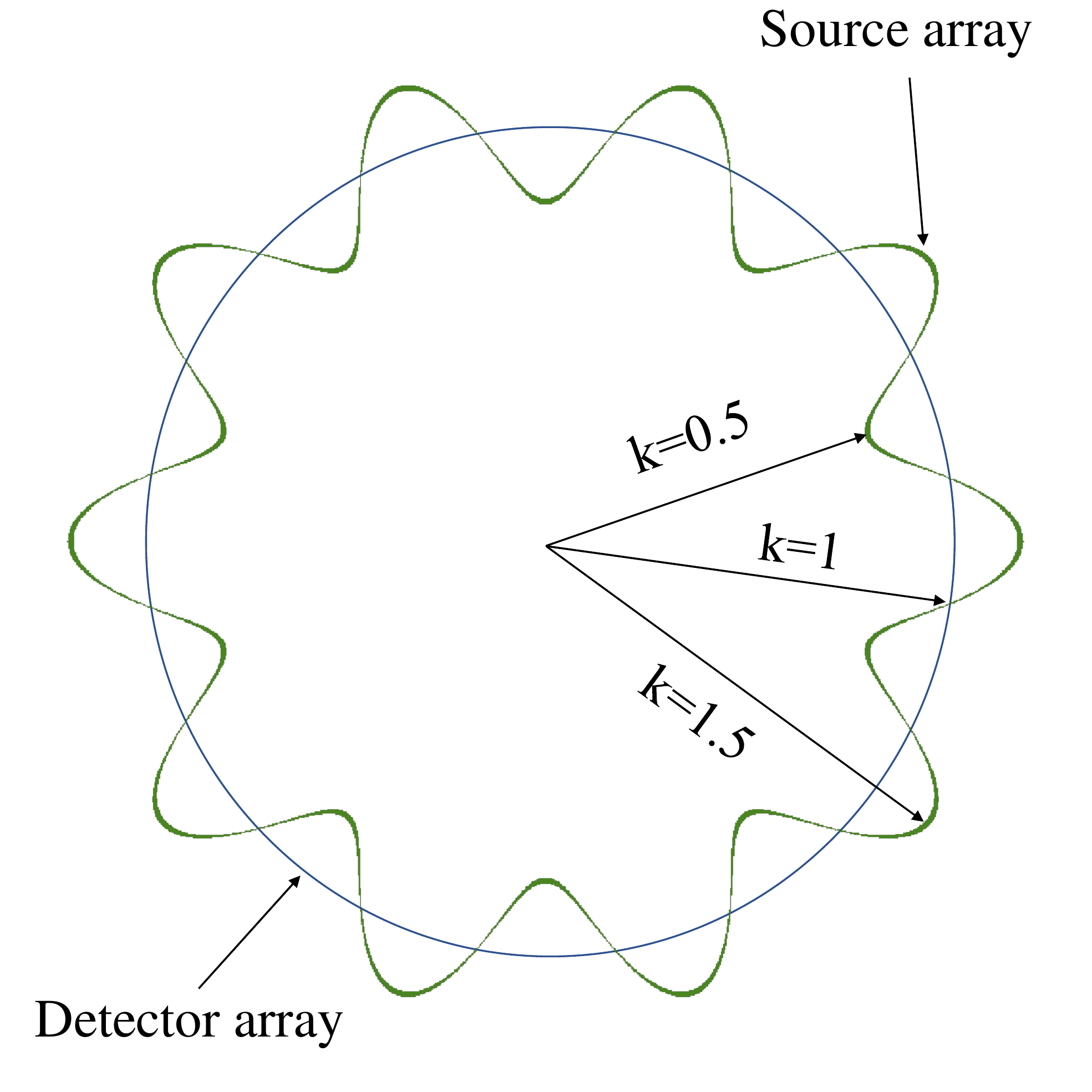}
	\caption{A schematic of GEGCT with dynamic NROD that might be useful for CT system design of new CT with distributed X-ray source in some special applications. }
	\label{fig:dk-1} 
\end{figure}
Apart from the GEGCT with fixed $k$, it would be interesting and practically useful to explore the GEGCT with dynamic $k$ as well, potentially beneficial to stationary CT system design. 
For instance, in the previous studies, stationary CT systems usually have a dual-ring or rectangle-shaped source and detector. \cite{wu2021stationary, tucker2020design, RN133, RN134}
GEGCT with dynamic $k$ could provide more freedom for positioning the X-ray source with more applicable trajectory as shown in Fig.~\ref{fig:dk-1}. Such an arrangement can accommodate more source points to improve the view sampling of stationary CT with distributed source, or allow more flexible positions of the source when scanning irregular objects.

It is seen that except for the two particular points of $k$ ($k=0$, and $k=1$) where conventional exact and shift-invariant FBP algorithm could be applied directly, GEGCT is non-equiangularly sampled.
In the potential applications of GEGCT with fixed or dynamic $k$, $0<k<1$ or $k>1$ is largely expected, how to get a practical but highly accurate shift-invariant FBP algorithm is worth taking careful considerations and is a major goal of this paper. 
\subsection{FBP algorithm with Shift-invariant Filtering}

The projection of GEGCT defined in the Eq.~(\ref{eqline}) has a similar expression as that of the traditional equiangular fan beam CT. So, taking the equiangular fan-beam FBP algorithm as reference, an FBP reconstruction formula of GEGCT with fixed $k$ can similarly be written as
\begin{equation}
	\begin{aligned}
		f(x,y) = \frac{1}{2}\int_{0}^{2\pi} d \beta
		{ \frac{1}{L^2} P_{f}(\alpha_0,\beta) }, \quad
		P_{f}(\alpha_0,\beta)  = 
		\int_{- \alpha_m}^{\alpha_m}  d \alpha
		{D\cos(\alpha)p(\alpha,\beta)h_{sin}(\alpha_0 - \alpha)},
	\end{aligned}
\end{equation}
where, $h_{sin}(t)$ is the ramp filter determined by
\begin{equation}
h_{sin}(t) = \int_{- \infty}^{+\infty} |\rho| e^{j2 \pi \rho *\sin t} d \rho,
\nonumber
\end{equation}
and $P_{f}(\alpha_0,\beta)$ is the projection after the filtering operation. In this formula, $\alpha_0$ and $L$ share the same definitions in Eq.~(\ref{eqline}).

In the standard equiangular fan-beam CT, $\alpha$ is uniformly sampled, which eventually makes the filtering operator shift-invariant.
For GEGCT, however, it is the fan angle with respect to the detector focus $\gamma$ that is uniformly sampled.
Taking the coordinate transfer in Eq.~(\ref{trans_akgama}), one can have
\begin{equation}
\mathrm{d}\alpha \mathrm{d}\beta 
= \alpha_k' (\gamma) \mathrm{d}\gamma \mathrm{d}\beta
= \frac{k \cos (\gamma) + 1 }{T^2(\gamma)} \mathrm{d}\gamma \mathrm{d}\beta,
\nonumber
\end{equation}
where
\begin{equation}
T(\gamma) = \sqrt{1+2k\cos(\gamma)+k^2}.
\nonumber
\end{equation}
Hence, the FBP reconstruction form for GEGCT with fixed $k$ can be re-writen as
\begin{equation}
	\begin{aligned}
		f(x,y) = \frac{1}{2} \int_{0}^{2\pi} \mathrm{d} \beta
		\frac{1}{L^2} P_f(\gamma_0,\beta)  , \,
		P_f(s_0,\beta)  = \int_{-{\gamma}_m}^{{\gamma}_m} \mathrm{d}\gamma
		p_k(\gamma,\beta) K(\gamma_0, \gamma) h_{sin}(\gamma_0 - \gamma)
		J(\gamma)
		\label{equa:2}
	\end{aligned}
\end{equation}
where
\begin{equation}
K(\gamma_0, \gamma) = \frac{\sin^2(\gamma_0 - \gamma)}{\sin^2(\alpha_k(\gamma_0) - \alpha_k(\gamma))}
\nonumber
\end{equation}
\begin{equation}
J(\gamma) = D \cos(\alpha_k(\gamma)) \alpha_k'(\gamma),
\nonumber
\end{equation}
and $\gamma_0$ is the fan angle with respect to the detector focus when $\alpha = \alpha_0$, i.e. 
\begin{equation}
\gamma_0 = \alpha^{-1}_{k}(\alpha_0).
\nonumber
\end{equation}

In the case of dynamic $k$, $k$ becomes a function of view angle $\beta$. Hence, the weighted FBP reconstruction needs to be adjusted accordingly. With $R$ and $\mathrm{DID}$ kept fixed (as shown in the Fig.~\ref{fig:dk-1}), we have
\begin{equation}
D(\beta) = \Delta D + R - \mathrm{DID} = k(\beta) R + R - \mathrm{DID} 
\nonumber
\end{equation}
Since $\alpha$ is a function of $k$, it changes with $\beta$ as well.
Therefore, for GEGCT with dynamic $k$,
\begin{equation}
(\theta,t) = (\alpha(\gamma,\beta)+\beta,D(\beta)\sin \alpha(\gamma,\beta)).
\nonumber
\end{equation}
The Jacobian for coordinate transfer will change to
\begin{equation}
\begin{aligned}
	J(\gamma,\beta) 
	&= \left(D\cos(\alpha)-D'(\beta)\sin(\alpha)\right) \alpha'(\gamma) \\
	&= \left(D\cos(\alpha)-Rk'(\beta)\sin(\alpha)\right) \alpha'(\gamma).
\end{aligned}
\nonumber
\end{equation}
By replacing $J(\gamma)$ in Eq.~(\ref{equa:2}) with $J(\gamma,\beta)$, we obtain the modified FBP reconstruction for GEGCT with dynamic $k$.

The weighting and filtering operations in the FBP algorithm represent the transform of $p_k(\gamma,\beta)$ to $P_f (\gamma_0,\beta)$.
In general, if such a transform can be written as a convolution, it can be efficiently implemented via the FFT technique.
Thus, in the FBP algorithm of GEGCT, shift-invariance is equivalent to finding a weighted convolution decomposition that satisfies \cite{RN181}
\begin{equation}
	K(\gamma_0, \gamma) = A(\gamma)B(\gamma_0-\gamma)C(\gamma_0).
\end{equation} 

By substituting Eq.~(\ref{trans_akgama}) into $K(\gamma_0, \gamma)$ of Eq.~(\ref{equa:2}), we can get
\begin{equation}
	K(\gamma_0, \gamma) = 
	T^2 (\gamma)T^2 (\gamma_0){\left(\frac{\cos(\frac{\gamma_0 - \gamma}{2})}{G(\gamma_0, \gamma)}\right)}^2 
	\label{equa:K}
\end{equation}
where
\begin{equation}
G(\gamma_0, \gamma) = \cos(\frac{\gamma_0 - \gamma}{2})+k\cos(\frac{\gamma_0 + \gamma}{2}).
\nonumber
\end{equation}
Therefore, developing a shift-invariant FBP formula can be realized equivalently looking for a weighted convolution decomposition satisfying
\begin{equation}
G(\gamma_0,\gamma) = A_1(\gamma)B_1(\gamma_0-\gamma)C_1(\gamma_0).
\nonumber
\end{equation}
In Appendix A, we prove that $G(\gamma_0, \gamma)$ cannot be exactly decomposed as $A(\gamma)B(\gamma_0-\gamma)C(\gamma_0)$ except for $k=0$ or $k=1$.
This suggests that we shall instead seek for approximate decomposition,
\begin{equation}
	A(\gamma)B(\gamma_0-\gamma)C(\gamma_0) \approx K(\gamma_0, \gamma).
\end{equation}
In the next sections, we will present three viable weighting strategies consisting of pre-filtering ($A(\gamma)$), filtering ($B(\gamma_0-\gamma)$) and post-filtering ($C(\gamma_0)$) weights, respectively.
\subsection{Weighting Strategy}
When deriving appropriate weights, theoretically the following criterion
\begin{equation}
\min_{A,B,C}\{ \!\! \iint_{\gamma,\gamma_0} \!\!\!\!\!\!
h^2 (\gamma_0 - \gamma) {(K(\gamma_0, \gamma)\!\! - \!\! A(\gamma)B(\gamma_0-\gamma)C(\gamma_0) )}^2  \,d \gamma  \,d \gamma_0 \}
\label{cr:1}
\tag{i}
\end{equation}
is helpful to get the optimal formula.
However, direct derivation based on Criterion (\ref{cr:1}) is complicated, less practical and hard to be solved. As a result, alternatively we focus on getting practical but highly precise weights intuitively or empirically. 
In this paper, we investigate at three different approaches with details as follows.

\subsubsection{Besson Weights}

If one rewrite the integral variables in Eq.~(\ref{equa:2}) using variables $(\gamma, \gamma_0)$ instead of $(\gamma, \beta)$, then in this dual variable space, the integral range of $\gamma$ will be from $-\gamma_m$ to $\gamma_m$ and integral range of $\gamma_0$ from $0$ to $\arcsin(\frac{\sqrt{x^2+y^2}}{D})$, which always includes $\gamma_0 = 0$. 
On the other hand, the filter kernel $h(\gamma_0-\gamma)$ reaches maximum value when $\gamma_0 = \gamma$.

Therefore, based on Criterion (\ref{cr:1}) above, one could come up a simple and natural approximation as

\begin{equation}
	\begin{aligned}
		A(\gamma)B(0)C(\gamma)  = K(\gamma, \gamma) = \frac{T^2(\gamma)T^2(\gamma)}{{(k \cos(\gamma)+1)}^2} , \,
		A(\gamma)B(-\gamma)C(0) = K(\gamma, 0) = T^2(\gamma).
	\end{aligned}
\end{equation}

With the symmetric property of $(\gamma_0,\gamma)$, let $A(x) = C(x)$ and $B(x)=B(-x)$,
and by using normalization condition $A(0) = 1$, the final weights could be expressed as
\begin{equation}
	\begin{aligned}
		A(\gamma) = C(\gamma) = \frac{1+2k \cos(\gamma)+k^2}{(k+1)(k \cos(\gamma)+1)} , \quad
		B(\gamma) = (k \cos(\gamma)+1)(k+1).
	\end{aligned}
\label{eq_B}
\end{equation}

When $k=0$ or $k=1$, $K(\gamma_0, \gamma) = A(\gamma)B(\gamma)C(\gamma)$, suggesting that this weight decomposition holds strictly.

It is worth noting that the weights derived above is consistent with weights that Besson established in 1996. When Besson studied fan-beam configurations that have exact shift invariant FBP reconstruction algorithm, he also provided an FBP approximate algorithm for the fourth generation CT where $k$ is close to $1$. Besson's approximation can be expressed as
\begin{equation}
{\sin}^2( \alpha_0 - \alpha (\gamma) ) \approx
\frac{\alpha' (\gamma_0)\alpha' (\gamma) {\sin}^2( \alpha (\gamma_0 - \gamma))}{\alpha' (\gamma_0 - \gamma)\alpha' (0)},
\end{equation}
where the corresponding pre-filtering, filtering and post-filtering weights were respectively written as
\begin{equation}
	\begin{aligned}
		A(\gamma) = C(\gamma)  = \frac{\alpha' (0)}{\alpha' (\gamma)} , \quad
		B(\gamma) = \frac{\sin^2(\gamma)\alpha' (\gamma)}{\sin^2(\alpha(\gamma))\alpha' (0)} 
	\end{aligned}
\end{equation}
They are consistent with Eq.~(\ref{eq_B}).
\subsubsection{Polynomial Weights}
From Eq.~(\ref{equa:K}), decomposition of $J(\gamma_0, \gamma)$ leads to the decomposition of $G(\gamma_0, \gamma)$,
\begin{equation}
	\begin{split}
	G(\gamma_0, \gamma) = \cos(\frac{\gamma_0 - \gamma}{2})+k \cos(\frac{\gamma_0 + \gamma}{2}) 
	\approx
	A_1 (\gamma) B_1 (\gamma_0-\gamma) C_1 (\gamma_0).
	\end{split}
\end{equation}

Taking advantages of the symmetric property of $(\gamma_0,\gamma)$ in $G(\gamma_0, \gamma)$, we have $A_1 (x) = C_1 (x)$, and $A_1$, $B_1$ and $C_1$ are all even functions. Different from the Besson weights, we can approximate $A_1$, $B_1$ and $C_1$ by just second-order polynomials, i.e.
\begin{equation}
	\begin{split}
		\frac{\cos(\frac{\gamma_0 - \gamma}{2})+k \cos(\frac{\gamma_0 + \gamma}{2})}{k+1} \approx
		(1+a \gamma^2) (1+a \gamma_0^2)
		(1+b(\gamma_0-\gamma)^2),
	\end{split}
\end{equation}
where, 
\begin{equation}
a = -\frac{k}{4k+4},\quad b = \frac{k-1}{8k+8},
\nonumber
\end{equation}
according to the derivation in Appendix B. As a result, by combining it with Eq.~(\ref{equa:K}), the final polynomial weights can be written as
\begin{equation}
	\begin{aligned}
		A(\gamma) = C(\gamma) = \frac{1+2k \cos(\gamma)+k^2}{(k+1)^2{(1+a\gamma^2)}^2} , \quad
		B(\gamma) = \left(\frac{cos(\gamma/2)(k+1)}{1+b\gamma^2}\right)^2
	\end{aligned}
\end{equation}
Certainly, $A_1$, $B_1$ and $C_1$ can be approximated by four or higher order polynomials as well. From the derivation in Appendix B, the coefficients of four-order terms are
\begin{equation}
a_4 = -\frac{k^2-2k}{96 {(k+1)}^2}, \quad b_4 = \frac{5k^2-6k+1}{384 {(k+1)}^2}.
\nonumber
\end{equation}

It is worth noting that theoretically the FBP reconstruction by using the polynomial weights will not be exact even when $k$ is $0$ or $1$.
\subsubsection{Empirical Weights}
Empirically, we also come up with a simple weighting formula that works with a relatively high accuracy as long as $k$ is close to 0 or 1.
\begin{equation}
	\begin{aligned}
		K(\gamma_0,\gamma)
		=
		\frac{\sin^2(\gamma_0 - \gamma)}{\sin^2(\alpha(\gamma_0) - \alpha(\gamma))} 
		\approx
		m_A^4 (\gamma) * \frac{\sin^2(\gamma_0 - \gamma)}{\sin^2(m_B(\gamma_0-\gamma))},
	\end{aligned}
\end{equation}
where
\begin{equation}
m_B = \int_{- \gamma_m}^{\gamma_m} \alpha_k'( \gamma ) \textrm{d} \gamma,
\quad
m_A (\gamma) = \frac{\gamma / \alpha}{\mathrm{mean}(\gamma / \alpha)}.
\nonumber
\end{equation}
In this empirically weighted FBP, post-filtering weighting is missing, leading to inferior performance in terms of reconstruction accuracy when compared with the other two weighting strategies. 
\subsubsection{Accuracy Analysis for the Weighted Strategies}
\begin{figure}[htb]
	\centering
	\includegraphics[width=0.6\textwidth]{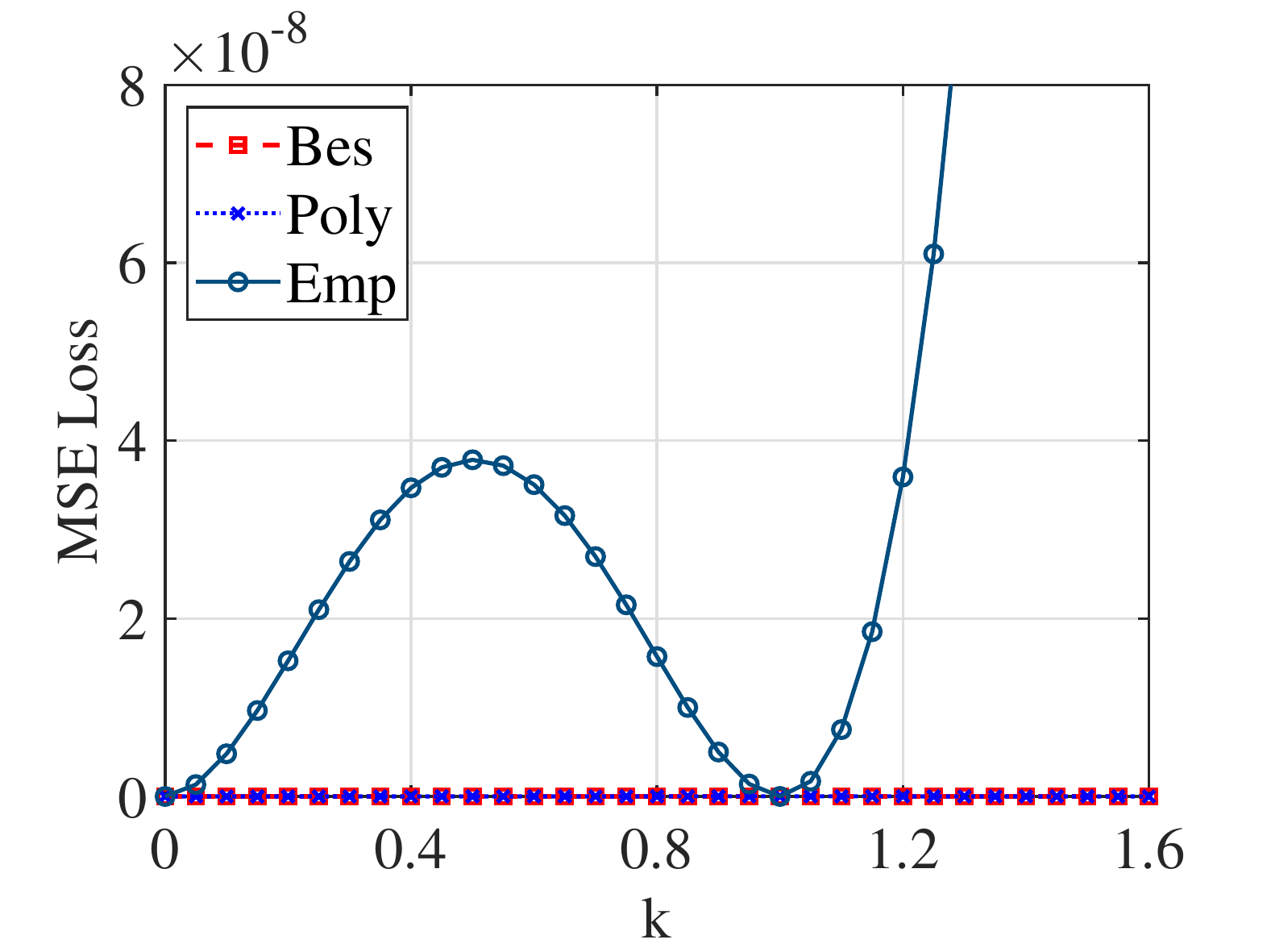}
	\caption
	{The theoretical error of approximate weights for different $k$. The MSE loss is computed basing on criterion (\ref{cr:1}) with semi fan angle of $20^{\circ}$ and interval of $0.01^{\circ}$.}
	\label{fig:pos} 
\end{figure}
In the cases of $k$ close to $0$ or $1$, the weighted FBP algorithms get highly accurate reconstruction as Besson has found in the literature. 
Under a wide range of $k$, whether the approaches still work effectively is vital for related CT system design and optimization. 
Taking the dual-ring of source and detector based stationary CT in Fig.~\ref{fig:applyCT2} for instance, the ratio of detector radius to source radius is the $k$. If the weighted FBP algorithms can work effectively under a wider range of $k$, more freedom and space can be allowed for the detector ring and source ring, respectively.

Although there is a difficulty for direct derivation based on Criterion (\ref{cr:1}), the mean square error (MSE) it represents, can be used to reveal the theoretical errors as $k$ changes.
As shown in Fig.~\ref{fig:pos}, for $k$ varying from 0 to 2, both the Besson and the polynomial weights have MSE losses less than $10^{-10}$; while the empirical weights has a relatively greater error as expected. 

\subsection{3D FDK-type Reconstruction}

For 3D cone-beam CT reconstruction, Feldkamp-Davis-Kress(FDK) algorithm, which was derived from FBP method, is used widely due to its simplicity and efficiency. \cite{CTFDK,FDK}
For GEGCT, the FDK-type 3D reconstruction formula for GEGCT could be expressed as:
\begin{equation}
	\begin{split}
		f(x,y,z) = \frac{1}{2}
		\int_{0}^{2 \pi} \mathrm{d} \beta
		\frac{1}{L^2} \int_{- {\gamma}_m}^{{\gamma}_m} \mathrm{d}\gamma 
		p_k(\alpha, \gamma, b(z))
		J \, \cos \eta K(\gamma_0, \gamma)
		h_{\sin}(\gamma_0 - \gamma).
	\end{split}
\end{equation}

the cone angle $\eta$ and the height of the virtual central detector elements $b(z)$ in FDK algorithm of GEGCT has different forms from that of the traditional FDK algorithm. Distance between source and detector elements of GEGCT is
$R_1 = \sqrt{R^2+D^2+2RD \cos (\gamma_0)}$. 
Therefore,
\begin{equation}
b(z) = \frac{z}{L}*R_1 ,\quad
\cos \eta = \frac{R_1}{\sqrt{R_1^2+b(z)^2}}.
\end{equation}

\begin{figure}[htb]
	\centering
	\begin{subfigure}[b]{0.45\textwidth}
		\centering
		\centerline{\includegraphics[width=\textwidth]{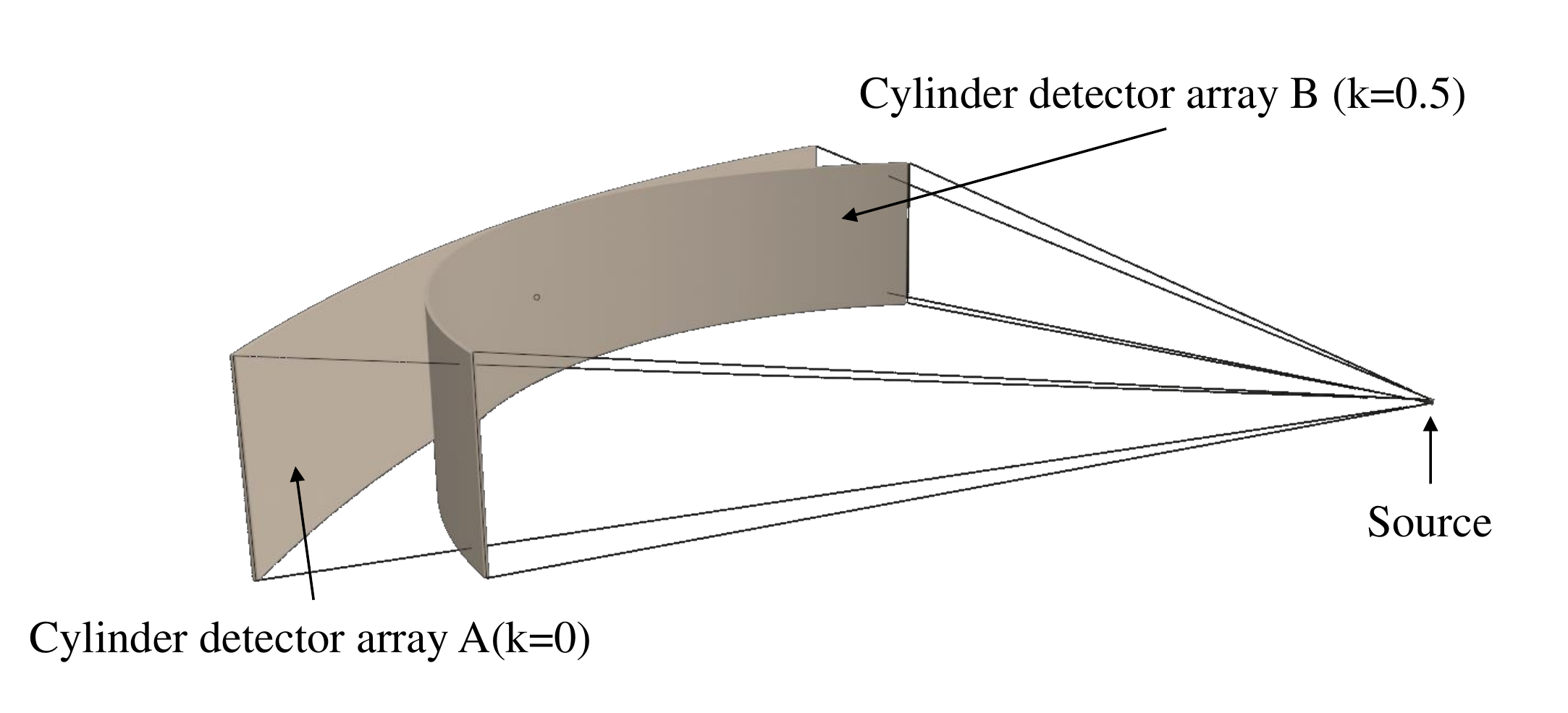}}
		\caption{}
		\label{fig:3d5-1}
	\end{subfigure}
	\hfil
	\begin{subfigure}[b]{0.45\textwidth}
		\centering
		\centerline{\includegraphics[width=\textwidth]{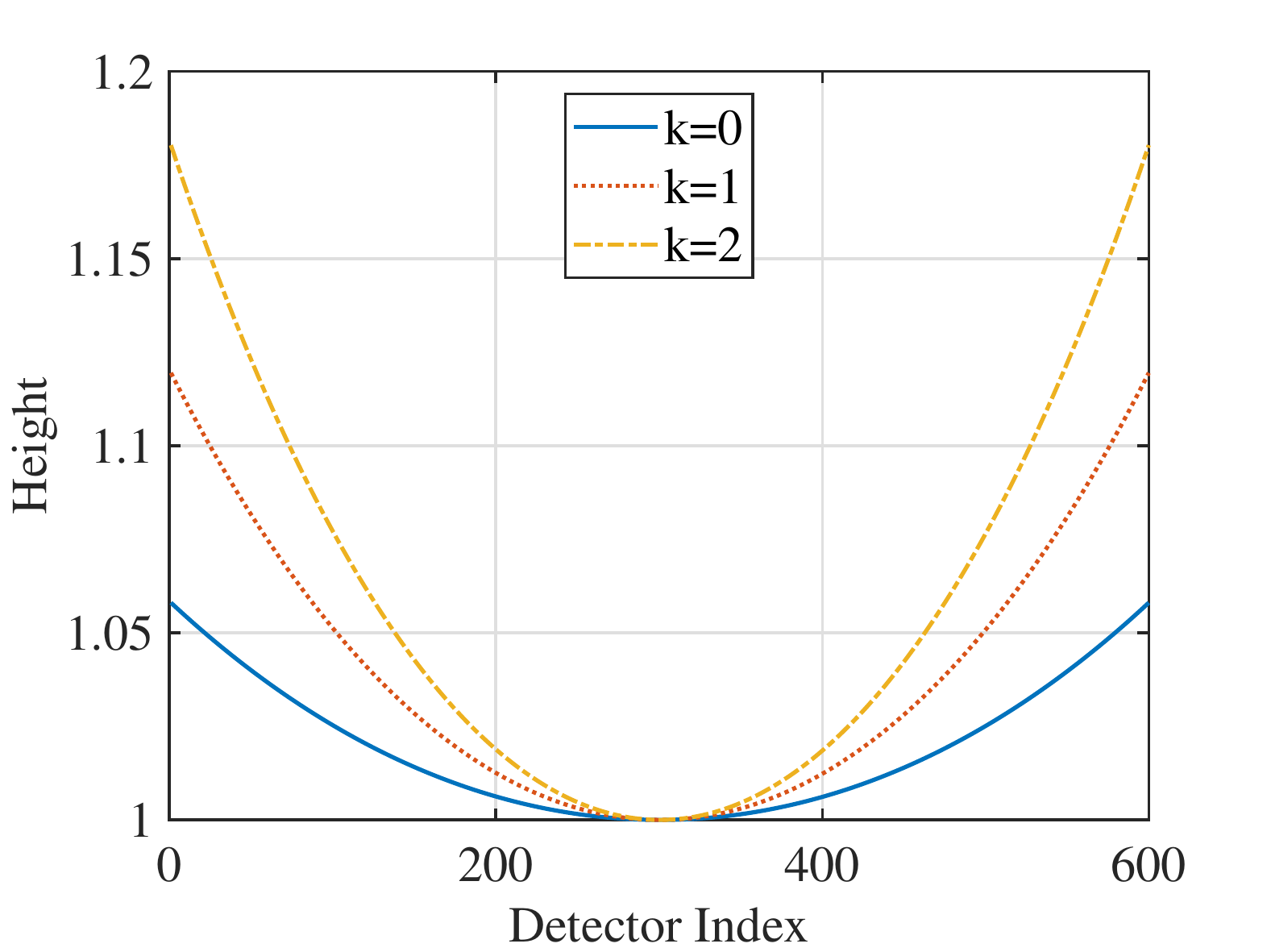}}
		\caption{}
		\label{fig:3d5-2}
	\end{subfigure}
	\caption{ The schematic of two cylindrical detectors with the same height but different $k$ (a), and the U-shaped curves of filtering direction on virtual detector plane as $k$ changes (b).}
	\label{fig:3dfdk}
\end{figure}

The value of $k$ will affect the cone beam artifact for GEGCT.
As illustrated in Fig.~\ref{fig:3d5-1}, when the length of the detector arc, DID and $D$ all keep the same, the curvature of the detector arc increases as $k$ increases. For the detector elements which are away from the mid-plane, it would also have bigger cone-angles as $k$ increases.
Fig.~\ref{fig:3d5-2} shows the shape of the projection from a detector row off-plane on the virtual detector plane with $D$ being 300 mm and $\mathrm{DID}$ being 600 mm. 
The different U-shaped curves indicate the cone beam artifact will be different as $k$ changes.

\section{Experiments and Results}

In order to analyze the performance of the presented weighted FBP algorithms for GEGCT with fixed or dynamic $k$, in terms of reconstruction accuracy and effectiveness, a comprehensive phantom simulation study along with a mimicked GEGCT scan of lung by using a clinical lung CT dataset was conducted in this paper, focusing on the reconstruction accuracy, spatial resolution, the impact of varying $k$, as well as its 3D performance where cone-beam artifact is of concern. The imaging parameters of GEGCT are listed in Table \ref{tab:CTscan}. It is noted that, change $D$ would change $k$ for GEGCT with $R$ and DID fixed. Peak Signal to Noise Ratio(PSNR) and Structural Similarity(SSIM) are selected as two major metrics for comparison.
In this paper, Bes is short for Besson weights, Poly for polynomial weights (the second-order term approximation) and Emp for empirical weights.

\begin{table}[h]
	\tabcolsep = 0.3 cm
	\renewcommand{\arraystretch}{1.1}
	\begin{center}
		\caption{Parameters of simulations 
			\label{tab:CTscan}}
			\resizebox{0.6\textwidth}{!}{
			\begin{threeparttable}
				\begin{tabular} {cc}
					\hline
					Parameters & Values \\
					\hline
					Detector radius ($R$) \tnote{*}   & 500 mm \\
					detector-isocenter distance (DID)\tnote{*}   & 500 mm \\
					Number of detector elements  & 1200 \\
					Interval of detector elements & 1 mm \\
					Size of reconstruction & 512*512 \\
					Pixel size of reconstruction & 1 mm * 1 mm\\
					Number of view angles & 1000 per rotation\\
					\hline
				\end{tabular}
				\begin{tablenotes}
					\footnotesize
					\item[*] It is the value in most cases unless otherwise indicated additionally.
				\end{tablenotes}
		\end{threeparttable}}
	\end{center}
\end{table}

\subsection{GEGCT with fixed NROD}
\subsubsection{Images Reconstructed by Different Weighted FBP Algorithms}
In order to evaluate performance of the presented three weighting strategies, we simulated GEGCT scans of Shepp-Logan phantom with $k$ being $2$ and $1.1$, respectively.
\begin{figure}[htb]
	\centering
	\includegraphics[width=0.7\textwidth]{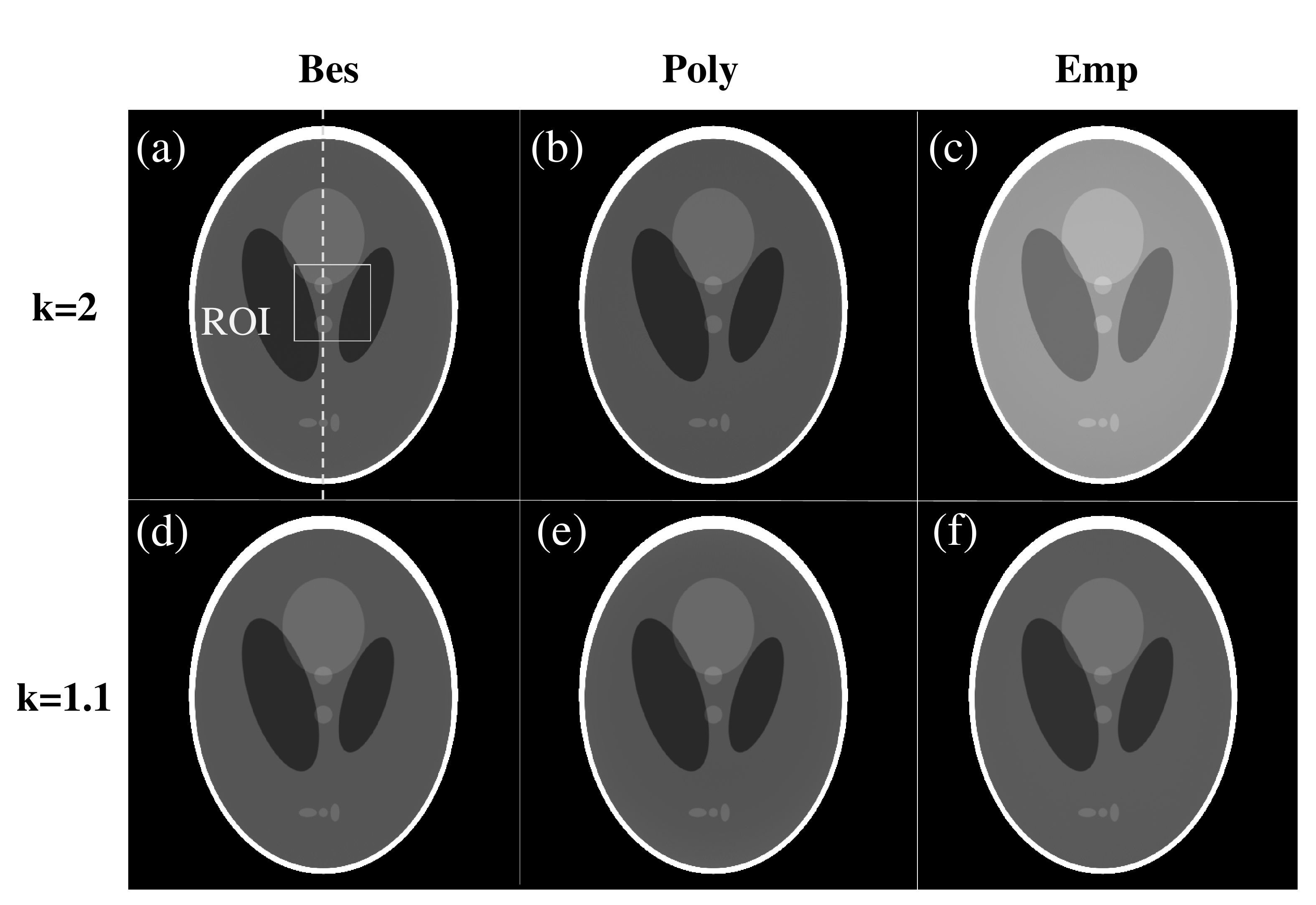}
	\caption{Reconstructions using different weighting FBP algorithms. $k$ is $2$ for (a)-(c); $k$ is $1.1$ for (d)-(f). The region of interest (ROI) labeled in (a) is selected to compute PSNR and SSIM for reconstruction accuracy comparison. Display window: [980,1100] HU.}
	\label{fig:SL}
\end{figure}

\begin{table}[h]
	\tabcolsep = 0.8 cm
	\renewcommand{\arraystretch}{0.9}
	\begin{center}
		\caption{PSNR and SSIM of CT images in the selected ROI in Fig.~\ref{fig:SL}.}
		\label{tab_SL}
		\resizebox{0.6\textwidth}{!}{
			\begin{threeparttable}
				\begin{tabular} {c|ccc}
					\hline
					k& Methods & PSNR(dB)&SSIM\\
					\hline
					1     &      &  41.07&0.985\\
					\hline
					& Bes   &39.81&0.979\\
					
					2     & Poly      &39.33&0.979\\
					
					& Emp      &  11.43&0.740\\
					\hline
					&  Bes      &  40.87&0.984\\
					
					1.1      & Poly     &  40.19&0.984\\
					
					&  Emp      &  32.09&0.979\\
					\hline
				\end{tabular}
		\end{threeparttable}}
	\end{center}
\end{table}

\begin{figure}[htb]
	\centering
	\begin{subfigure}[b]{0.48\textwidth}
		\centering
		\centerline{\includegraphics[width=\textwidth]{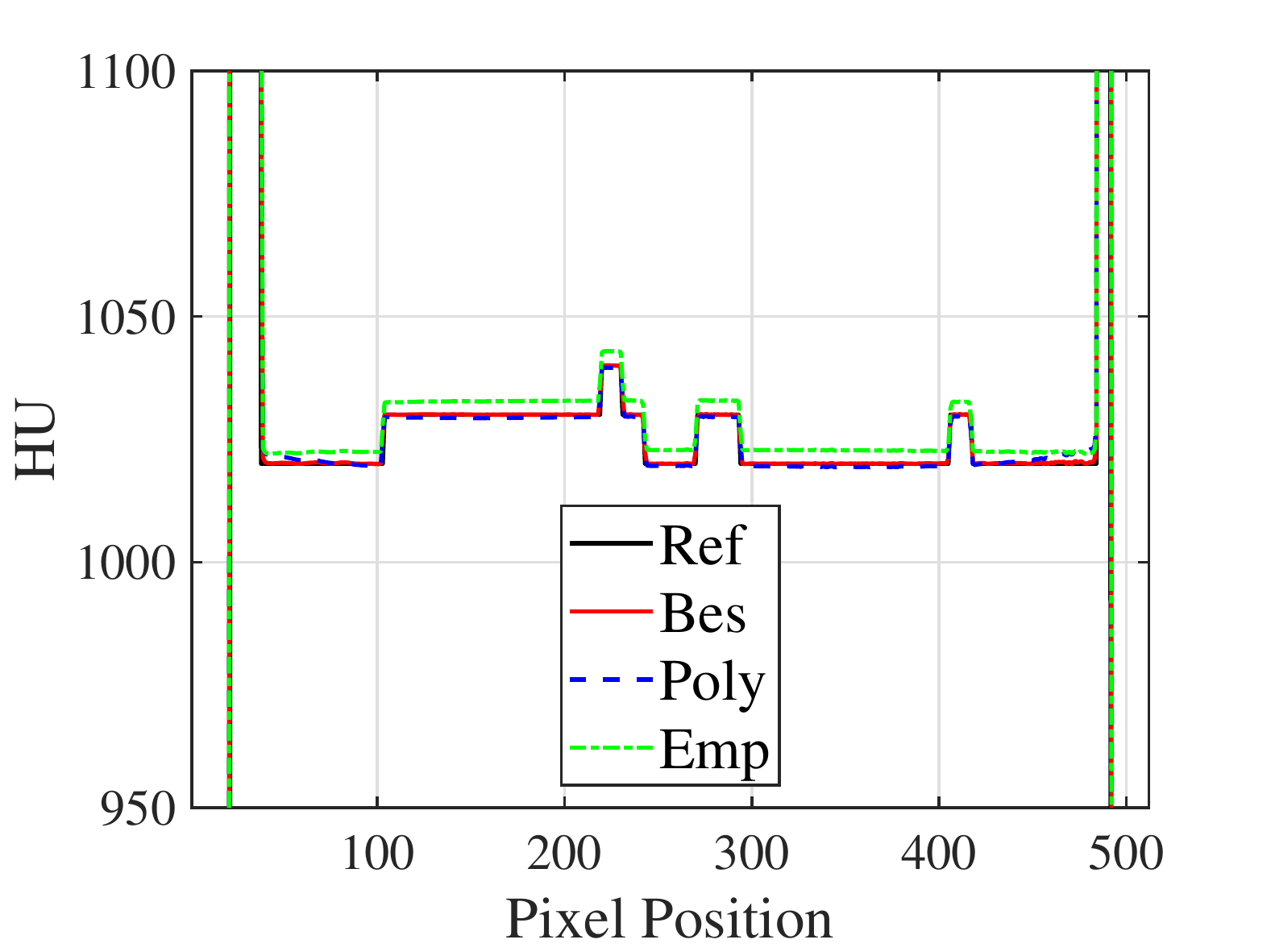}}
		\caption{$k =1.1$}
		\label{fig:slp1}
	\end{subfigure}
	\hfil
	\begin{subfigure}[b]{0.48\textwidth}
		\centering
		\centerline{\includegraphics[width=\textwidth]{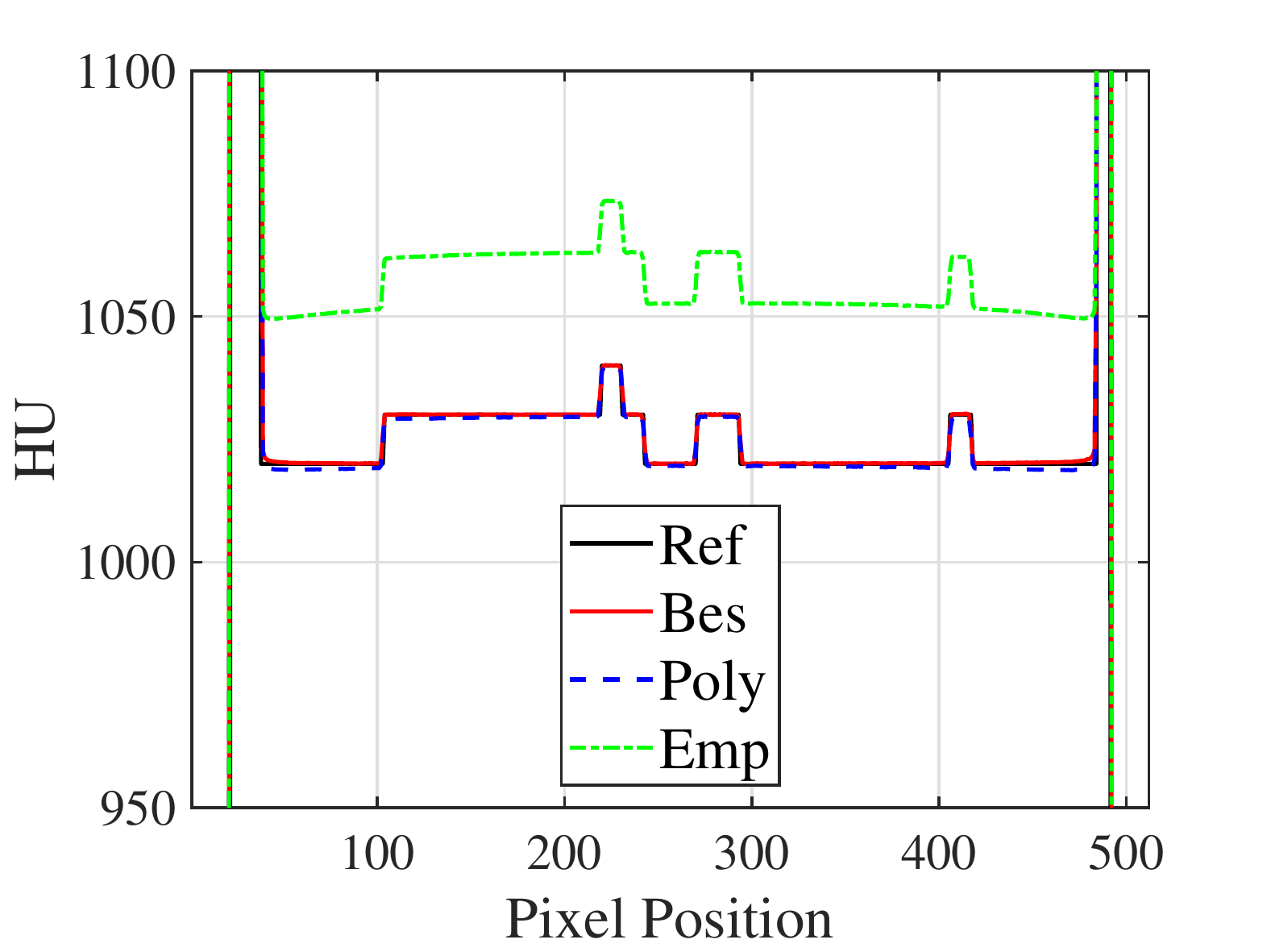}}
		\caption{$k = 2$}
		\label{fig:slp2}
	\end{subfigure}
	\caption{The profile labeled in Fig.~\ref{fig:SL} (a) for different weighted methods in the case of  $k = 1.1$ and $k = 2$, respectively. Ref is the profile of original phantom.}
	\label{fig:slp}
\end{figure}

Images reconstructed by using the three different weighted FBP formulas are shown in Fig.~\ref{fig:SL}. First, it is observed that empirical weights have relatively lower accuracy, especially when $k$ is off from 1, which is consistent with the theoretical analysis in Section II.C.4.
For empirical weights, it achieves lower PSNR and SSIM as summarized in Table \ref{tab_SL} and relatively greater error than the other two strategies as shown in Fig.~\ref{fig:slp}.
PSNR and SSIM of the other two weighted FBP algorithms are at the same high level as that of $k = 1$ (equiangular fan-beam CT).
The profiles in Fig.~\ref{fig:slp} show that Besson weights get the most accurate reconstruction while polynomial weights still have about $1$ hounsfield unit (HU) of error. Empirical weights have relatively lower accuracy.

\begin{figure}[htb]
	\centering
	\includegraphics[width=0.6\textwidth]{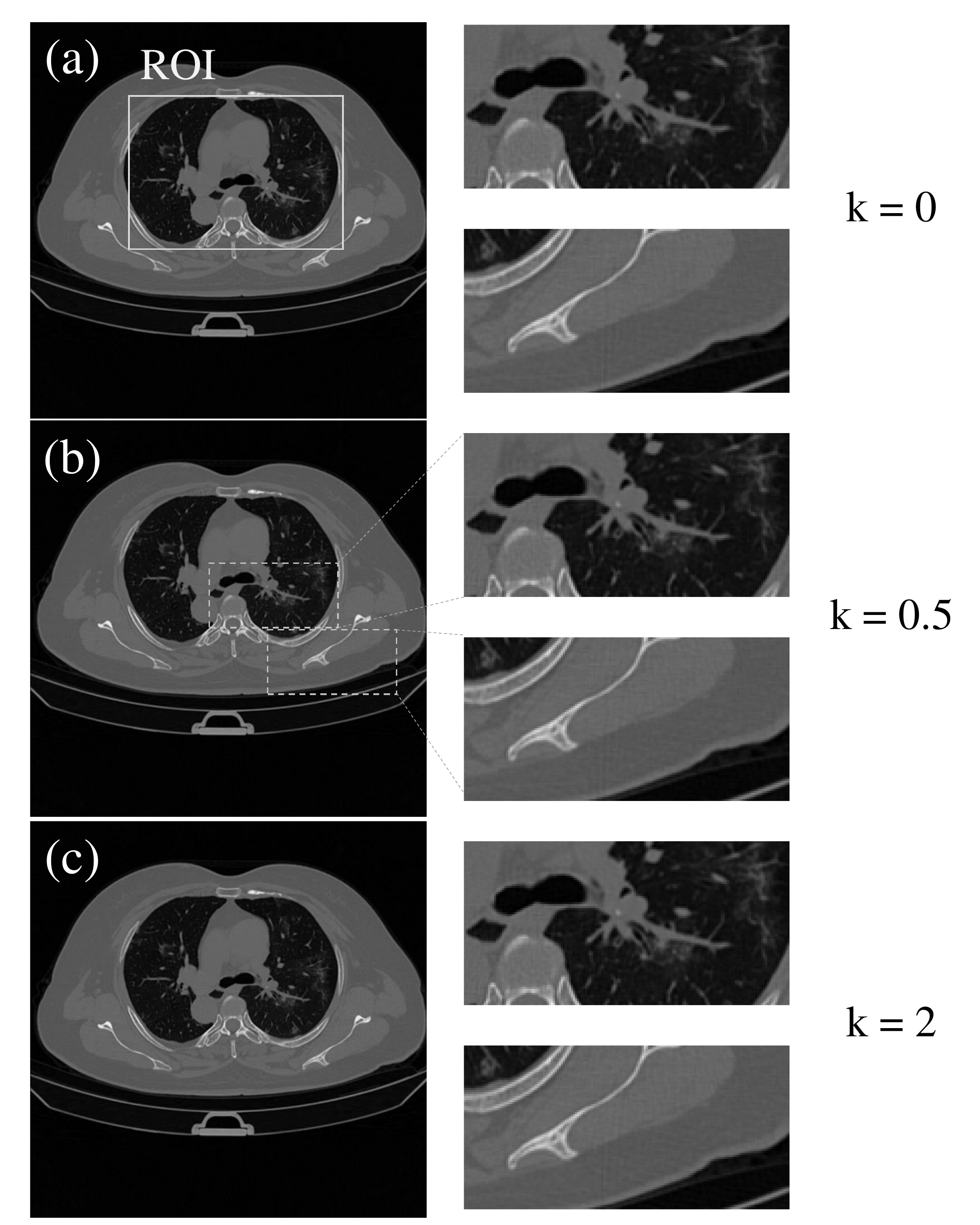}
	\caption{GEGCT reconstructions of lung CT data for different $k$, taking results of the polynomial weights as examples. Gray window: [0,1].}
	\label{fig:realslice}
\end{figure}

The effectiveness of the proposed weights is validated on real CT slices for a wide range of $k$ by mimicking a GEGCT scan using clinical lung CT dataset. \cite{jun2020covid}
In order to get the same geometric magnification ratio in the experiments, $D$ is fixed at $1000$ mm and $\mathrm{DID}$ is fixed at $500$ mm while $R$ varies. 
Taking polynomial weights as examples in Fig.~\ref{fig:realslice}, there is no obvious difference among $k$ = 0,  0.5 and 2. The zoomed-in images in Fig.~\ref{fig:realslice} show the same level of crisp detail of texture.
As listed in Table \ref{tab:CTslice}, in the selected ROI, the HU accuracy of the reconstruction by using the proposed polynomial weights is consistent with that of the standard equiangular fan-beam CT.

\begin{table}[htb]
	\tabcolsep = 0.2 cm
	\renewcommand{\arraystretch}{1.1}
	\begin{center}
		\caption{PSNR and SSIM of CT images in the selected ROI for different $k$ in Fig.~\ref{fig:realslice}.}
		\label{tab:CTslice}
		\resizebox{0.48\textwidth}{!}{
			\begin{threeparttable}
				\begin{tabular} {c|ccccccc}
					\hline
					k   & 0\tnote{*} & 0.5 & 1 & 1.5 & 2 \\
					\hline
					PSNR(dB) & 37.59&	37.60  & 37.59&	37.59&	37.59 \\
					SSIM   & 0.979	&0.979	&0.979 &0.979	&0.979\\
					\hline
				\end{tabular}
				\begin{tablenotes}
					\footnotesize
					\item[*] The traditional equiangular fan-beam CT
				\end{tablenotes}
		\end{threeparttable}}
	\end{center}
\end{table}

\subsubsection{Applicable Range of NROD}
To clearly observe the performance of the proposed weights as NROD($k$) changes and find out the applicable range for $k$, we reconstructed a simple phantom of water cylinder, at a series of $k$ (0, 0.2, 0.5, 0.7, 1, 1.5, 2). In this simulation, the detector radius $R$ was set as $500$ mm for $k \geq 0.7$; and 900 mm for $k \leq 0.7$, to ensure a reasonable scan filed of view.
\begin{figure}[htb]
	\centering
	\begin{subfigure}{0.45\textwidth}
		\centering
		\includegraphics[width=\textwidth]{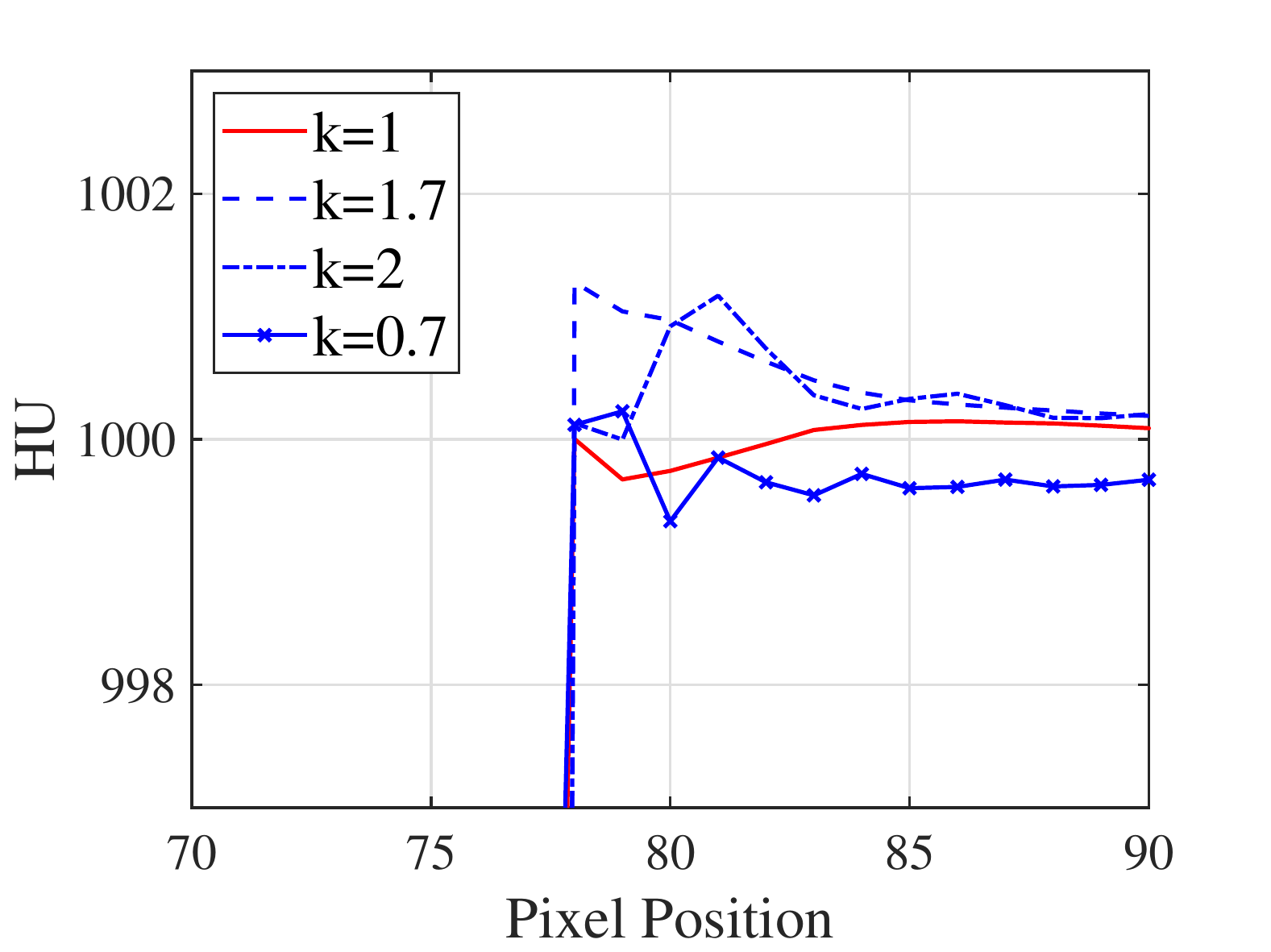}
		\caption{}
		\label{fig:water1}
	\end{subfigure}
	\begin{subfigure}{0.45\textwidth}
		\centering
		\includegraphics[width=\textwidth]{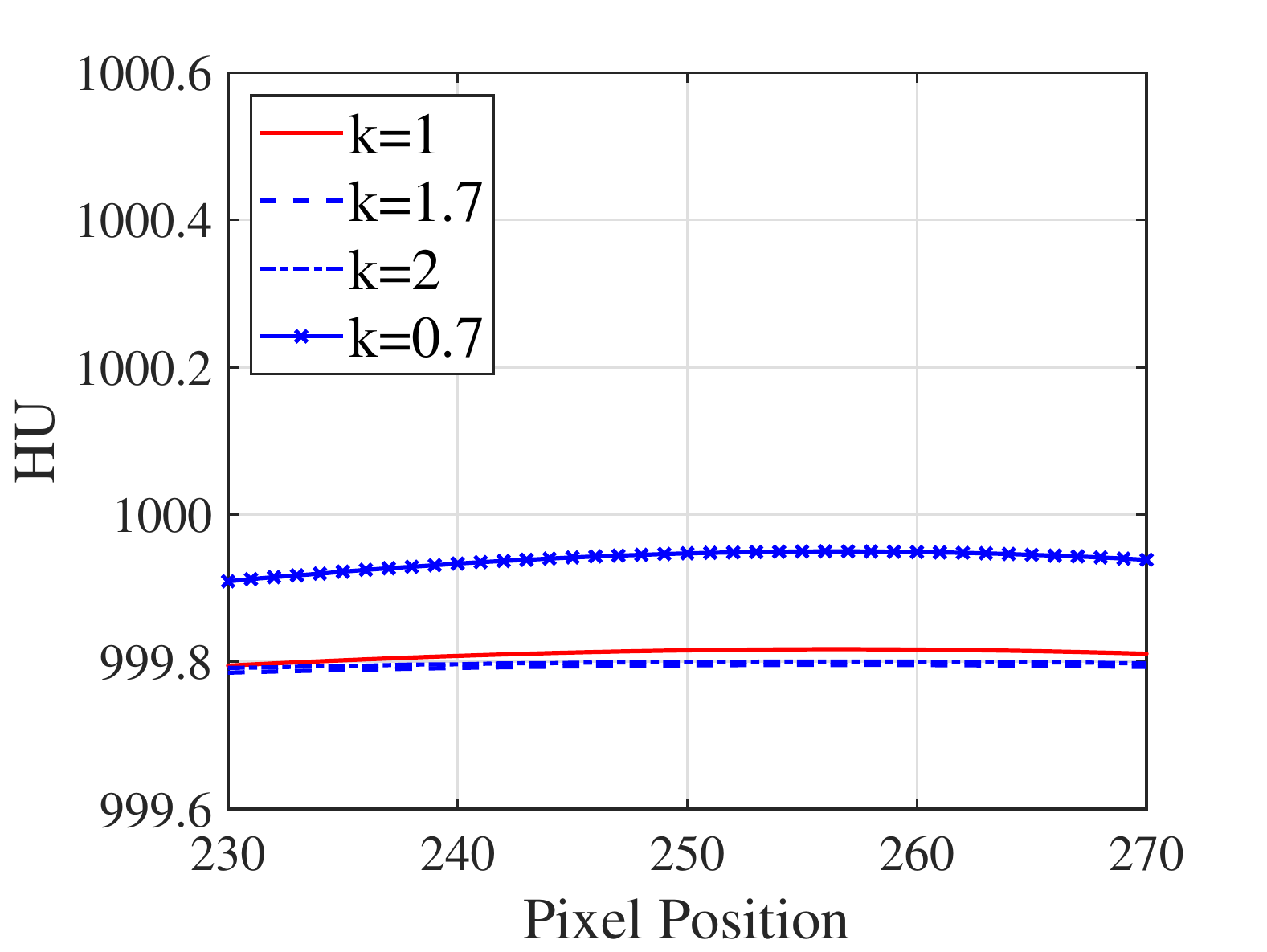}
		\caption{}
		\label{fig:water2}
	\end{subfigure}
	\begin{subfigure}{0.45\textwidth}
		\centering
		\includegraphics[width=\textwidth]{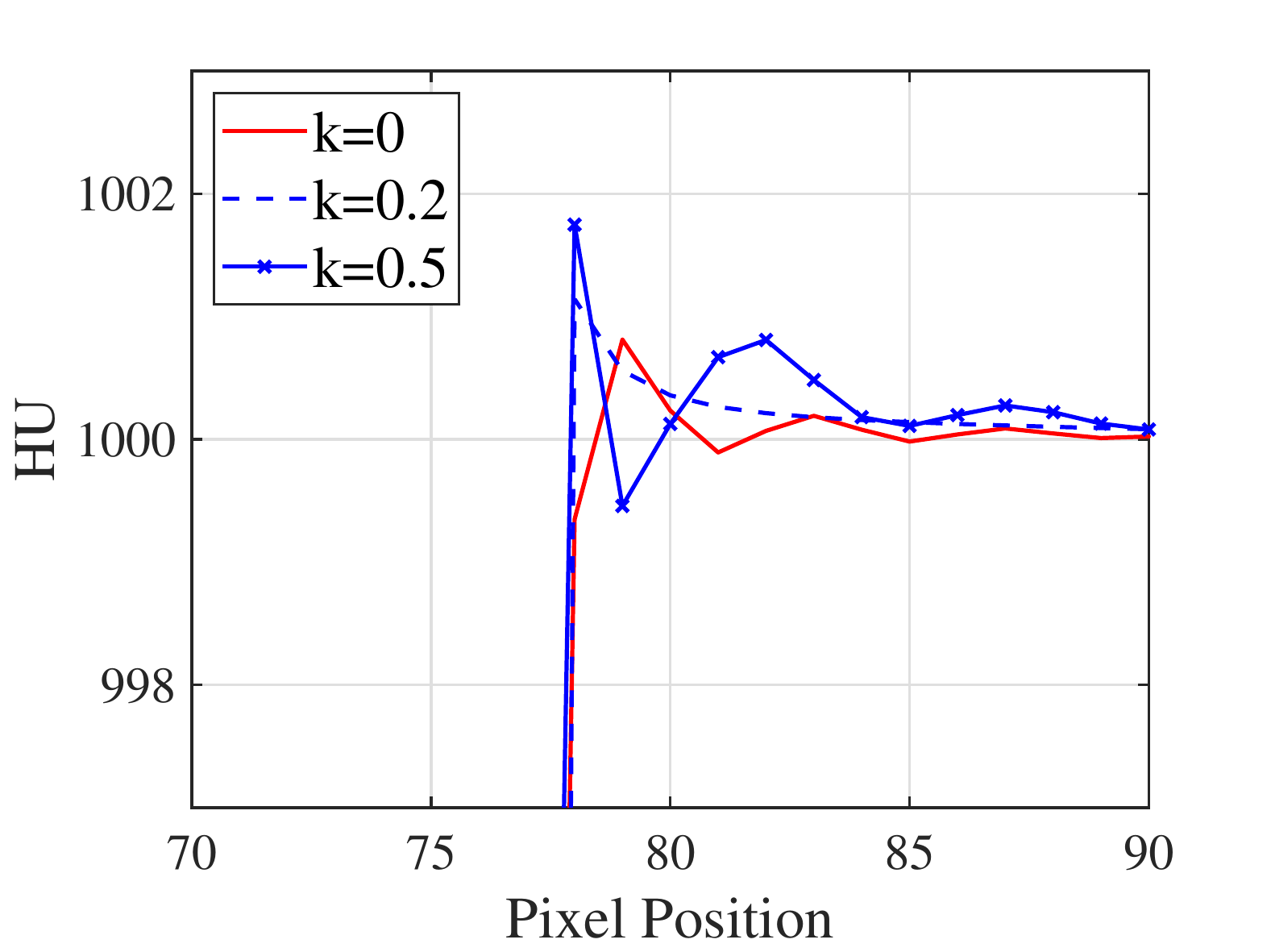}
		\caption{}
		\label{fig:water3}
	\end{subfigure}
	\begin{subfigure}{0.45\textwidth}
		\centering
		\includegraphics[width=\textwidth]{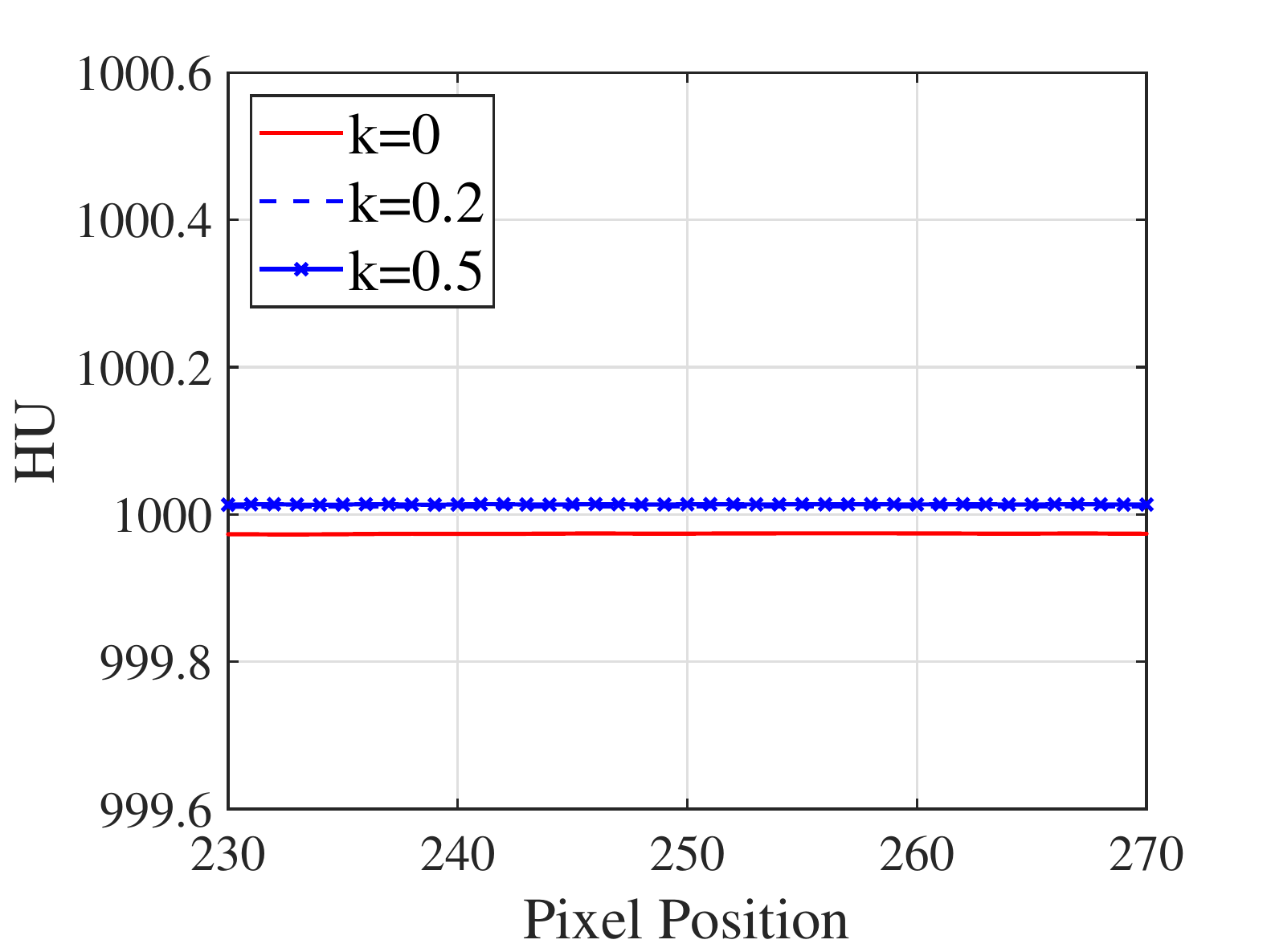}
		\caption{}
		\label{fig:water4}
	\end{subfigure}
	\caption
	{The central horizontal profiles of reconstructed CT images of water cylinder by using the Besson weights and polynomial weights with different $k$. (a) and (c) show the edges of water cylinders using the Besson weights; (b) and (d) show the center of water cylinders using the polynomial weights.}
	\label{fig:water} 
\end{figure}

The central horizontal profiles of reconstructed CT images of water cylinder in Fig.~\ref{fig:water} show that for GEGCT with a wide range of $k$ (from 0 to 2), the proposed FBP algorithm can achieve highly accurate reconstruction that is comparable to that of $k$ = 0 (the standard equiangular fan-beam CT).
The edge of water phantom has about $1$ HU of oscillation due to the Gibbs effect. For the CT number of water being $1000$ HU, the maximum errors of reconstruction are all less than $0.03\%$ in the center.
\subsection{GEGCT with dynamic NORD}
\begin{figure}[htb]
	\centering
	\includegraphics[width=0.65\textwidth]{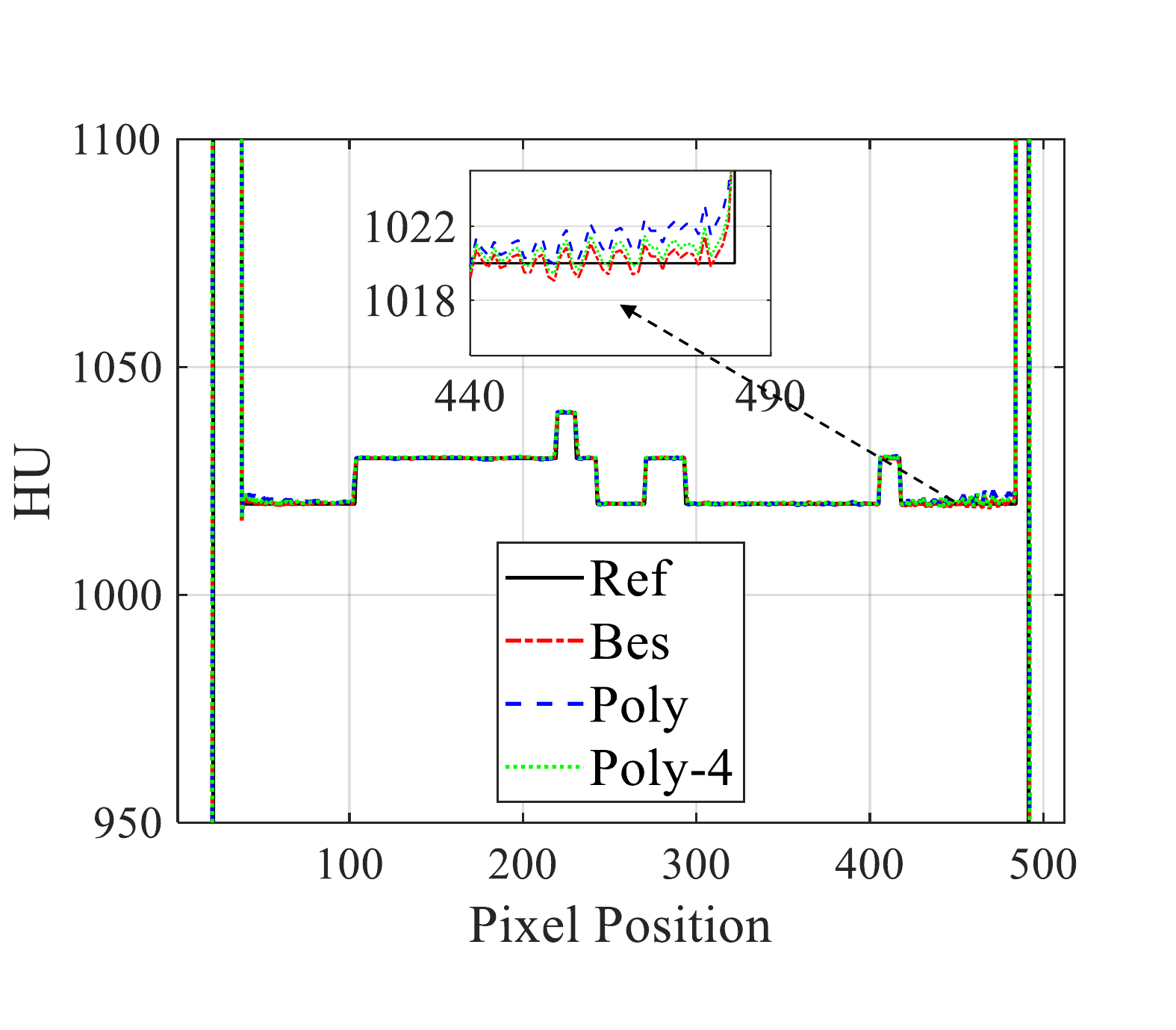}
	\caption{The profiles of reconstructions from a GEGCT scan of Shepp-Logan phantom with dynamic $k$, whose position is shown in the Fig.~\ref{fig:SL}. Poly-4 is the short for the reconstruction using the fourth-order polynomial weights. Ref is the profile of original phantom.}
	\label{fig:dk} 
\end{figure}

In this study, a GEGCT scan with dynamic NROD($k$) was simulated, where $k$, as a function of $\beta$, changes a lot during the scan (as shown in Fig.~\ref{fig:dk-1}),
\begin{equation}
k(\beta) = \frac{\Delta D}{R} = \frac{cos(m \beta)}{n} + 1
\nonumber
\end{equation}
Here, $m$ and $n$ are parameters related to the distributions of the source positions. In this paper, we choose $m = 8$ and $n = 2$ ($0.5\leq k \leq 1.5$) to validate our algorithms.

To compare the accuracy of GEGCT with dynamic $k$ with that of GEGCT with fixed one, a Shepp-Logan phantom was first simulated, and the detector radius $R$ was set as $610$ mm for $0.5 \leq k \leq 1.5$ to ensure a reasonable scan filed of view.
As shown in Fig.~\ref{fig:dk}, there exist some noticeable errors at the edge of high contrast section for polynomial weights. However, it is much improved using the fourth-order polynomial weights. And it is seen that the overall reconstruction accuracy is not affected by the use of dynamic $k$. The PSNR and SSIM of the selected ROI using second-order polynomial weights(shown in the Fig.~\ref{fig:SL}) are 42.31 dB and 0.9908 respectively, which is the same level as $k=1$.
\begin{figure}[htb]
	\centering
	\begin{subfigure}[b]{0.4\textwidth}
	\centering
	\includegraphics[width=\textwidth]{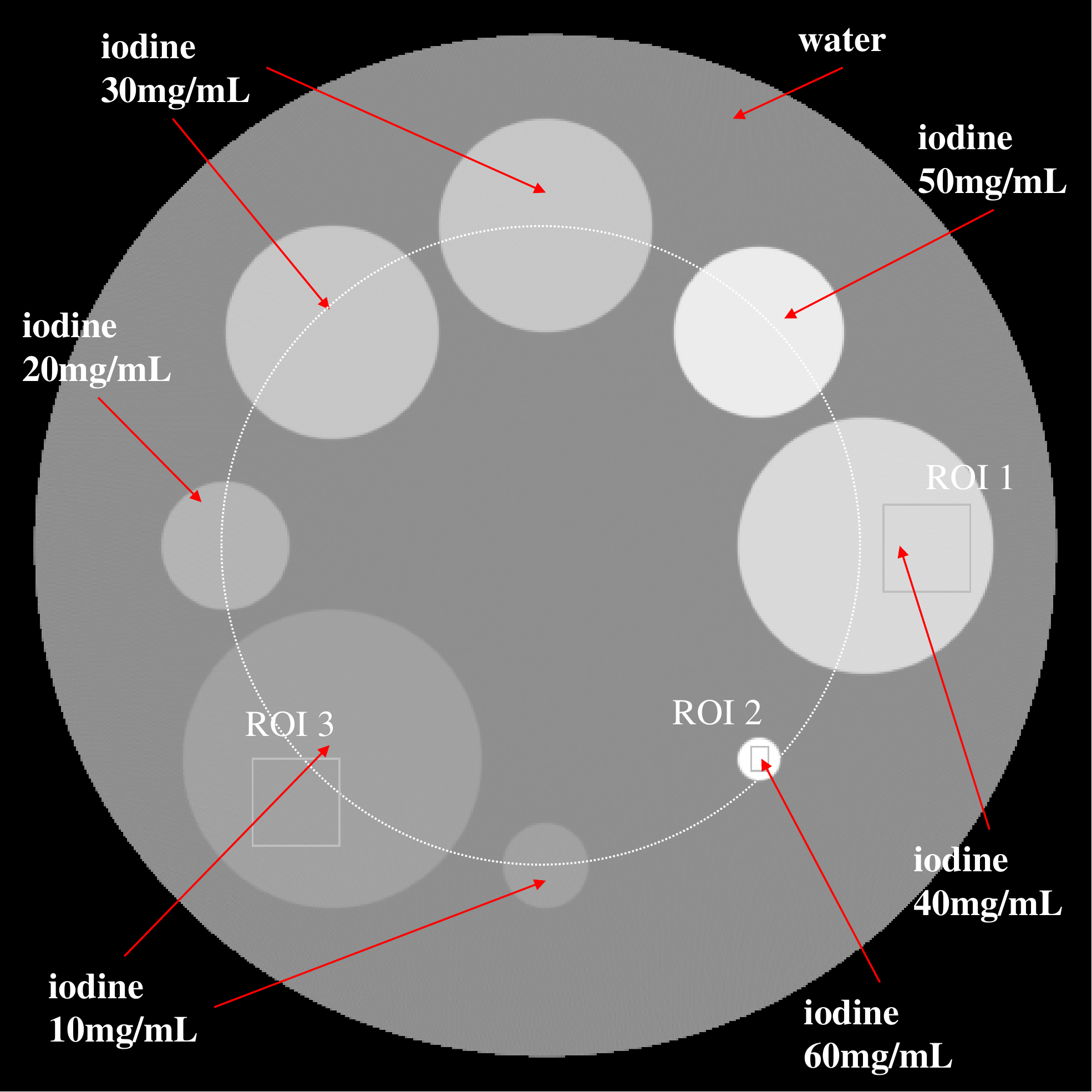}
	\caption{}
	\label{fig:varykimg}
	\end{subfigure}
	\begin{subfigure}[b]{0.5\textwidth}
	\centering
	\includegraphics[width=\textwidth]{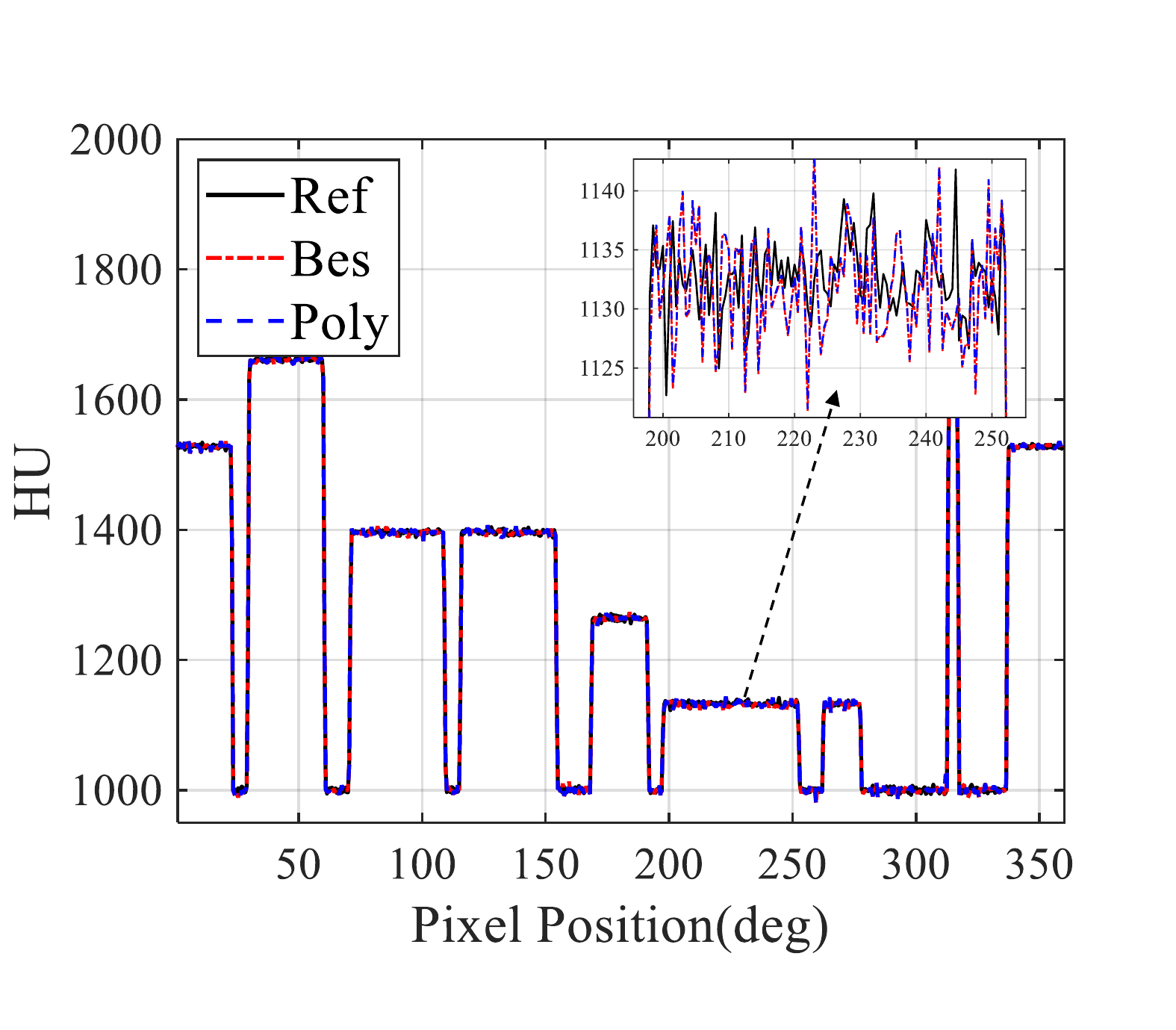}
	\caption{}
	\label{fig:dk5}
	\end{subfigure}
	\caption
	{(a) A digital phantom of 240-mm-radius water cylinder with eight iodine inserts. The ROIs labeled are selected to calculate a root mean square error. (b)The circular profiles of reconstructions from GEGCT scan of the digital phantom with dynamic $k$. Ref is result of GEGCT with fixed $k=1$.}
	\label{fig:12} 
\end{figure}

For the Besson weights and polynomial weights, in order to further validate the accuracy of the reconstruction in GEGCT with dynamic $k$, we further simulate a 240-mm-radius water cylinder with different densities of iodine inserts as shown in Fig.~\ref{fig:varykimg}. Again, the detector radius $R$ was set as $610$ mm.
Here, a root mean square error (RMSE) of CT numbers is used as metrics for measuring the reconstruction accuracy,
\begin{equation}
RMSE = 
\sqrt {\frac{1}{N} \sum_{i=1}^{N} {(\mu_i - \tilde{\mu_i})}^2 }
\nonumber
\end{equation}
where $\mu_i$ is the CT number (HU) in the selected ROI of reconstruction and $\tilde{\mu_i}$ is the corresponding reference CT number.

The profiles in Fig.~\ref{fig:dk5} show that both Besson weights and polynomial weights can get almost accurate reconstruction. The zoomed-in sub-image in Fig.~\ref{fig:dk5} illustrates that the weighted methods for GEGCT with dynamic $k$ have a little more volatility than that of GEGCT with fixed $k$. 
As listed in Table \ref{tab:rmse}, in the selected ROIs of the high-contrast phantom, RMSE for the GEGCT with dynamic $k$ are higher than that of GEGCT with fixed $k$.
\begin{table}[htb]
	\tabcolsep = 0.5 cm
	\renewcommand{\arraystretch}{1.1}
	\begin{center}
		\caption{RMSE of CT images in the selected ROIs for different methods in Fig.~\ref{fig:varykimg}.}
		\label{tab:rmse}
		\resizebox{0.45\textwidth}{!}{
			\begin{threeparttable}
				\begin{tabular} {c|ccc}
					\hline
					& Ref\tnote{*} & Bes & Poly \\
					\hline
					ROI 1 & 6.52    &9.16   &9.17\\
					ROI 2 & 5.33	&6.02	&6.03 \\
					ROI 3 & 4.24	&8.17	&8.21 \\
					\hline
				\end{tabular}
				\begin{tablenotes}
					\footnotesize
					\item[*] Ref is for GEGCT with fixed $k=1$, which also means the equiangular fan-beam CT.
				\end{tablenotes}
		\end{threeparttable}}
	\end{center}
\end{table}

\subsection{Spatial Resolution}
\begin{figure}[htb]
	\centering
	\includegraphics[width=0.6\textwidth]{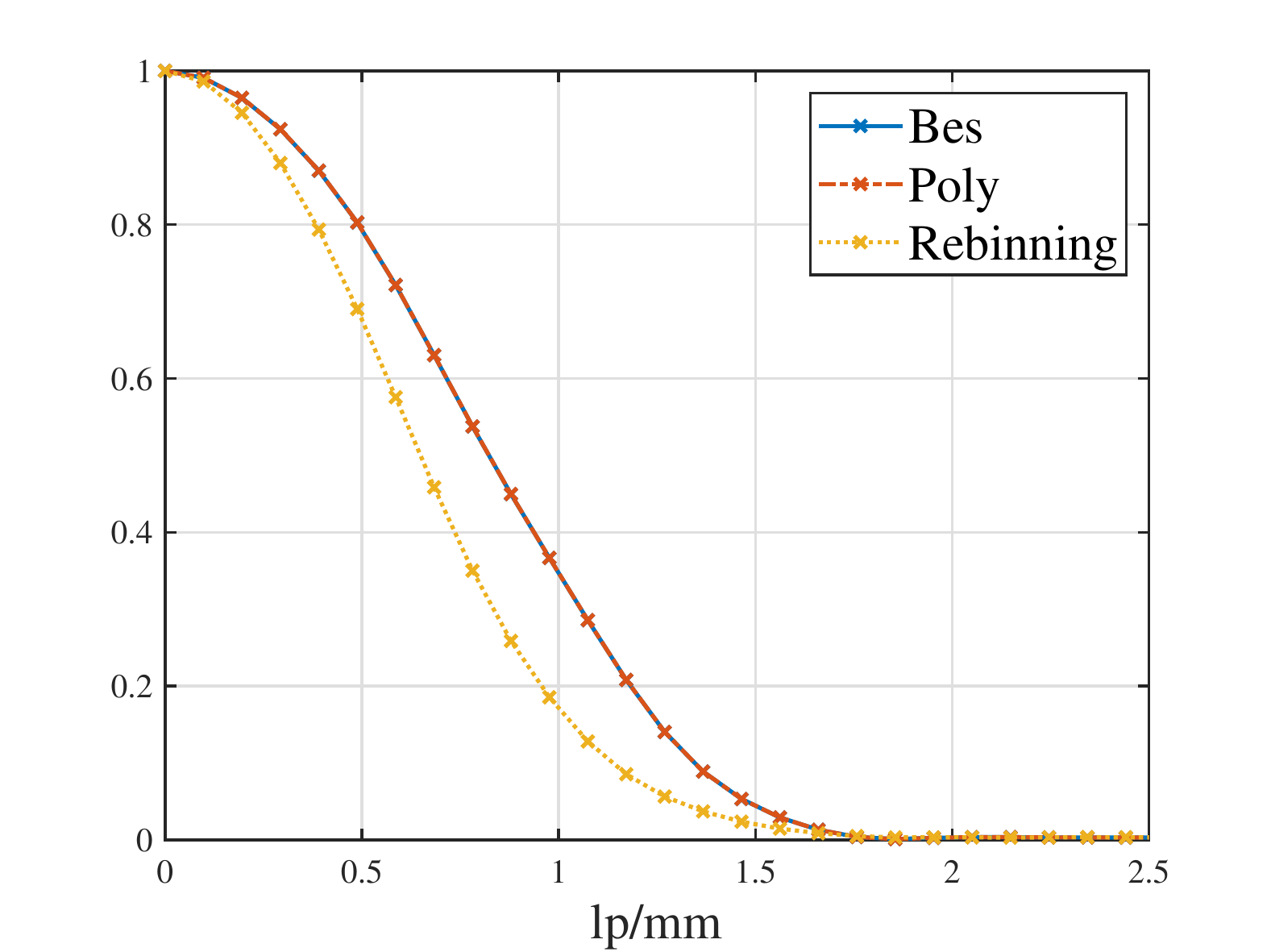}
	\caption{The modulation transfer function(MTF) curves from a GEGCT scan of a tungsten wire for $k=0.8$, reconstructed by using the direct FBP methods with Besson and polynomial weights, and the rebinning-to-equiangular method. The two weighted method achieve $1.35$ lp/mm of spatial resolution at $10\%$ MTF point, while the rebinning method only reach at $1.14$ lp/mm.}
	\label{fig:reso}
\end{figure}
To evaluate the performance of GEGCT reconstruction in terms of spatial resolution, we simulated a GEGCT scan of a tungsten wire with a diameter of $ 4 \mu m$  at $k = 0.8$, which is placed 100 mm away from the center.
The Modulation Transfer Function (MTF) curves are plotted along radial direction in Fig.~\ref{fig:reso}, for reconstructions by our proposed weighted FBP methods with no rebinning and by the rebinning-to-equiangular method, respectively.
As expected, the weighted FBP methods without rebinning achieve better spatial resolution with 1.35 lp/mm at $10\%$ MTF, while interpolation during rebinning-to-equiangular step costs about $16\%$ loss of MTF.

\subsection{3D Reconstruction}
To observe the 3D reconstruction performances from a GEGCT scan using the proposed weighted FBP algorithms, we carried out a simulation using the modified Defrise disk phantom which is commonly used to assess the cone-beam artifact. \cite{li2011} This phantom consists of five identically flattened ellipsoids in a ball of 120 mm in radius. The half-axis lengths of each ellipsoid are 60 mm, 60 mm, and 10 mm, positioned 40 mm apart in the Z direction. The CT number of the disks and ball are 2000 and 1000 HU, respectively.
During the GEGCT scan, $D$ is set as 300 mm while $\mathrm{DID}$ is 600 mm. A cylindrical detector is used with 600 × 500 detector elements, each having a size of 1 mm × 1 mm. 3D CT images are reconstructed with a size of 256 × 256 × 400 voxels. 
\begin{figure}
	\centering
	\begin{subfigure}[b]{0.45\textwidth}
		\centering
		\includegraphics[width=\textwidth]{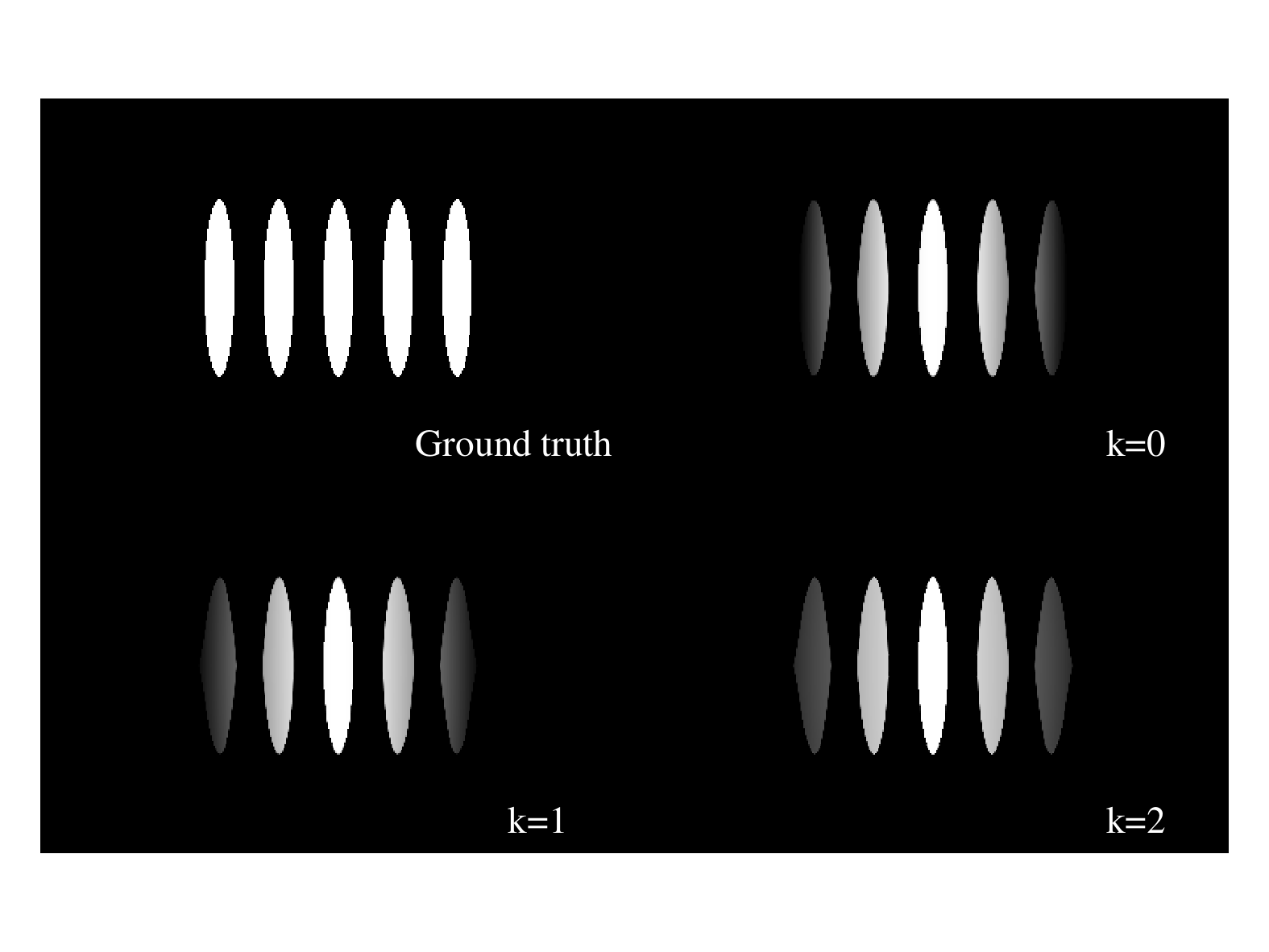}
		\caption{}
		\label{fig:3d2-1}
	\end{subfigure}
	\begin{subfigure}[b]{0.5\textwidth}
		\centering
		\includegraphics[width=\textwidth]{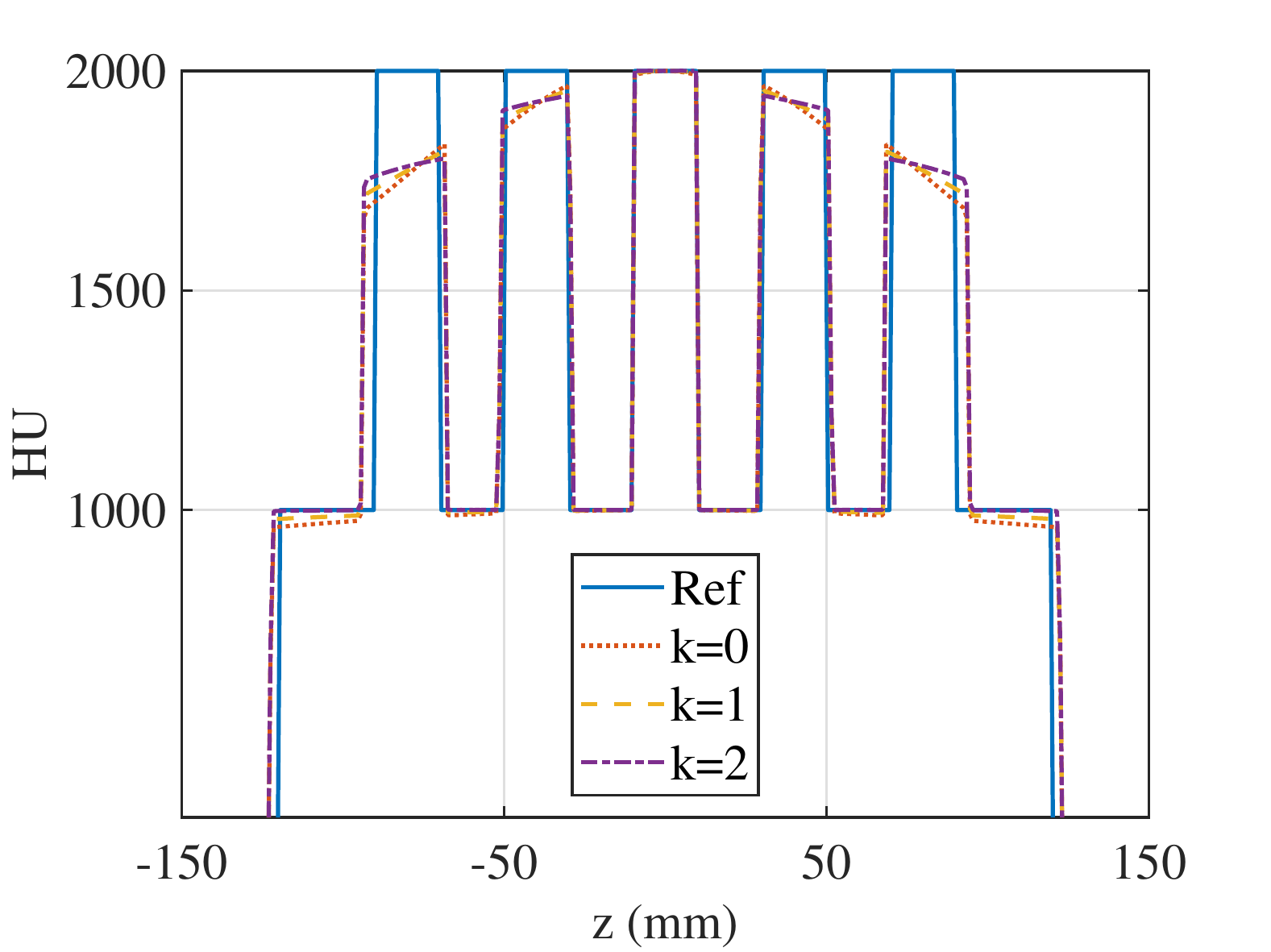}
		\caption{}
		\label{fig:3d2-2}
	\end{subfigure}
	\caption{3D reconstructions of modified Defrise disc phantom by using the polynomial weights. (a) The reformatted CT images in X-Z plane; display window: [1700,2000] HU, and (b) the central vertical profiles of the reformatted CT images. Ref is the profile of original phantom. }
	\label{fig:3d2}
\end{figure}
Fig.~\ref{fig:3d2} shows reformatted CT images in the XZ plane in the cases of $k$=0, 1, and 2, where the central vertical profiles are also plotted. It is quite obvious that larger $k$ leads to weaker cone beam artifacts.
\section{Discussion and Conclusions}

In this work, we proposed the concept of GEGCT and presented three approximate but shift-invariant FBP algorithms. 
A theoretical accuracy analysis and a comprehensive phantom simulation study along with a mimicked GEGCT scan of lung were conducted, validating the effectiveness of the weighted FBP algorithms in GEGCT with fixed or dynamic $k$. 
Among them, the Besson weights and our proposed polynomial weights can achieve highly accurate reconstruction at the same level as the standard equiangular fan-beam CT.
The quantitative comparison in terms of PSNR, SSIM, and spatial resolution measured by MTF on a variety of scanning object, demonstrated advantages and robustness of the presented FBP reconstruction methods. 
The overall performance of using the empirical weighting strategy with no post-filtering weight is inferior to the Besson and polynomial ones, suggesting that for GEGCT, post-filtering weighting is critical for a good approximate of shift-invariant FBP reconstruction as well.

This concept and algorithm study for GEGCT lay out a solid foundation for future GEGCT related system design and optimization.  
It is concluded that,for NROD ranging from $0.5$ to $1.5$, GEGCT with dynamic NROD could get highly accurate reconstructions, further suggesting that a varying source position radially during rotation is acceptable in terms of FBP reconstruction. 
Also, we revealed that less cone beam artifact is expected as NROD increases.

It is worth noting that for GEGCT reconstruction, rebinning to parallel-beam or uniformly sampled equiangular fan-beam is doable, but there might be a possible sacrifice of spatial resolution loss due to projection data interpolation as shown in Fig.~\ref{fig:reso}.
One can also choose iterative reconstruction approaches as well, which is more flexible to CT geometry but is usually more time consuming and more sensitive to data truncation. \cite{RN119}
Therefore, it is always beneficial to develop a direct and shift-invariant FBP algorithm for GEGCT, where the appropriate filtering and data weighting strategy is found to be effective and essential.

In this work, we focus more on the develop of practical reconstruction algorithms for GEGCT, while ignoring most of its physics and system aspects. It is noted that for GEGCT, due to the use of ring or arc-based detector whose focus is no longer on the source, the X-ray detection efficiency might be a little lower than that of the standard equiangular CT. Such an imaging geometry also makes the anti-scatter grid more difficult to deploy. These physics challenges are quite commonly encountered in stationary CT. As a matter of factor, GEGCT with dynamic NROD could be a good compromise for stationary CT with ring- or arc-shaped detector and source, as it provides a new way of maximizing the number of source points in a probably crowded space.

Future work includes development of a GEGCT prototype and applying the presented shift-invariant FBP algorithms into more GEGCT scenarios, as well as further exploring and evaluating its feasibility, robustness, and advantages. 
\section*{Acknowledgments}
This project was supported in part by grants from the National Natural Science Foundation of China (No. U20A20169 and No. 12075130).

\newpage
\appendix
\section{Proof of Nonexistence of Shift-invariant Property in general for $K(\gamma_0, \gamma)$ in Eq.~\ref{equa:K}}

Eq.~\ref{equa:K} tells us that the task becomes how to write
\begin{equation}
G(\gamma_0, \gamma) = \cos(\frac{\gamma_0 - \gamma}{2})+k\cos(\frac{\gamma_0 + \gamma}{2})
\label{eq:a1}
\end{equation}
as $A_1(\gamma)B_1(\gamma_0-\gamma)C_1(\gamma_0)$. 
First, we can add the condition $A_1(0)=1$, $C_1(0)=1$ and $B_1(0)=k+1$ to limiting the three functions. If $G(\gamma_0,\gamma) = A_1(\gamma)B_1(\gamma_0-\gamma)C_1(\gamma_0)$ holds in general, we have
\begin{equation}
\begin{aligned}
G(x, 0) &= A_1(x)B_1(x)C_1(0) = (k+1)\cos(\frac{x}{2}) \\
G(0, -x) &= A_1(0)B_1(x)C_1(-x) = (k+1)\cos(\frac{x}{2}) \\
G(x, -x) &= A_1(x) B_1(2x) C_1(-x)= k+\cos(x) \\
G(x,x) &= A_1(x)B_1(0)C_1(x) = k\cos(x)+1.
\end{aligned}
\nonumber
\end{equation}
Combining the parity of $A_1$, $B_1$ and $C_1$, we could get
\begin{equation}
\begin{aligned}
	& A_1(x) = C_1(x) = \sqrt{\frac{k\cos(x)+1}{k+1}}
	\\
	& B_1(x) = (k+1)\frac{\cos(\frac{x}{2})+k}{k \cos(\frac{x}{2})+1}.
\end{aligned}
\nonumber
\end{equation}
So the $G(\gamma_0,\gamma) $ can be denoted as
\begin{equation}
\begin{aligned}
G(\gamma_0,\gamma)  = ({k\cos(\gamma)+1}) ({k\cos(\gamma_0)+1}) \frac{\cos(\frac{\gamma_0-\gamma}{2})+k}{k \cos(\frac{\gamma_0-\gamma}{2})+1}
\end{aligned}
\label{eq:a2}
\end{equation}

Eq.~(\ref{eq:a1}) and Eq.~(\ref{eq:a2}) are not equal in general. $K(\gamma_0, \gamma)$ does not have the exact weighted decomposition for vary $k$. Hence, GEGCT's FBP reconstruction method doesn't have shift invariant property for $k$ generally.

\section{Derivation of The Second-order Polynomial weighting}
For the even function property of $G(\gamma_0, \gamma)$, we approximate $G(\gamma_0, \gamma)$ using the polynomials without odd-order terms,
\begin{equation}
	\begin{split}
		\frac{\cos(\frac{\gamma_0 - \gamma}{2})+k \cos(\frac{\gamma_0 + \gamma}{2})}{k+1}\approx
		(1+a_2 \gamma^2 + a_4 \gamma^4)
		(1+a_2 \gamma_0^2 + a_4 \gamma_0^4)(1+b_2 (\gamma_0-\gamma)^2 + b_4 (\gamma_0-\gamma)^4),
	\end{split}
\end{equation}
where, coefficients $a_2, a_4$ and $b_2, b_4$ are $k$ related. Let $\gamma=\gamma_0$ and $\gamma=-\gamma_0$, and we get
\begin{equation}
\begin{aligned}
	\frac{1+k \cos(\gamma)}{k+1}
	\approx &
	(1+a_2 \gamma^2 + a_4 \gamma^4)(1+a_2 \gamma^2 + a_4 \gamma^4)
	\\
	\frac{k+ \cos(\gamma)}{k+1}
	\approx &
	(1+a_2 \gamma^2 + a_4 \gamma^4)(1+a_2 \gamma^2 + a_4 \gamma^4)
	(1+4b_2 \gamma^2 + 16b_4 \gamma^4).
\end{aligned}
\nonumber
\end{equation}
Using Taylor Expansion,
\begin{equation}
\begin{aligned}
\sqrt {\frac{1+k \cos(\gamma)}{k+1}} 
&= 1 - \frac{k}{4k+4}\gamma^2 -\frac{k^2 - 2k}{96k^2+192k+96}\gamma^4
\cdots \\
\frac{k+ \cos(\gamma)}{1+k \cos(\gamma)}
&= 1 + \frac{k-1}{2k+2}\gamma^2 + \frac{5k^2-6k+1}{24k^2+48k+24}\gamma^4
\cdots ,
\end{aligned}
\nonumber
\end{equation}
we get
\begin{equation}
a_2 = -\frac{k}{4k+4}, \quad b_2 = \frac{k-1}{8k+8}.
\nonumber
\end{equation}
And the coefficients of four-order terms are
\begin{equation}
a_4 = -\frac{k^2-2k}{96 {(k+1)}^2}, \quad b_4 = \frac{5k^2-6k+1}{384 {(k+1)}^2}.
\nonumber
\end{equation}
Due to the use of truncated Taylor Expansion, FBP algorithm with the derived polynomial weights will not be theoretically exact even in the case of $k$ = 0 or 1. Polynomials of fourth order or higher are expected to be more accurate.
\newpage
\bibliographystyle{unsrt} 
\bibliography{Ref} 
\end{document}